\documentclass[%
 reprint,
 amsmath,amssymb,
 aps,
]{revtex4-2}

\usepackage{graphicx}
\usepackage{dcolumn}
\usepackage{bm}
\usepackage[capitalize]{cleveref}

\usepackage[dvipsnames]{xcolor}
\usepackage{tikz}

\renewcommand*\vec[1]{\mathbf{\bm{#1}}}
\begin{document}


\title{Active string fluids and gels formed by dipolar  active Brownian particles in 3D
}

\author{Maria Kelidou}
 \altaffiliation[Also at ]{Technical University Dortmund, Department of Physics, Dortmund, Germany}
 \author{Mohammad Fazelzadeh}
 \affiliation{Institute of Physics,  University of Amsterdam, Amsterdam, The Netherlands
}%
 \author{Baptiste Parage}
 \affiliation{Institute of Physics,  University of Amsterdam, Amsterdam, The Netherlands
}%
\author{Sara Jabbari-Farouji}%
 \email{s.jabbarifarouji@uva.nl}
\affiliation{Institute of Physics,  University of Amsterdam, Amsterdam, The Netherlands
}%

\date{\today}

\begin{abstract}
Self-propelled particles possessing  permanent magnetic dipole moments occur naturally in magnetotactic bacteria and in man-made systems like active colloids or micro-robots. Yet, the interplay between self-propulsion and anisotropic dipole-dipole interactions on dynamic self-assembly in three dimensions (3D) remains  poorly understood. We conduct Brownian dynamics simulations of active dipolar particles in 3D, focusing on the low-density regime, where dipolar hard spheres tend to form chain-like aggregates and percolated networks with increasing dipolar coupling strength. We find that strong active forces override dipolar attractions, effectively inhibiting chain-like aggregation and network formation.
 Conversely, activating particles with low to moderate forces results in a fluid composed of active chains and rings. At strong dipolar coupling strengths, this active fluid transitions into an active gel, consisting  of a percolated network of active chains. Although the overall structure of the active gel remains interconnected, the network experiences more frequent configurational rearrangements due to the reduced bond lifetime of active dipolar particles. Consequently, particles exhibit enhanced translational and rotational diffusion within the active fluid of strings and  active gels compared to their passive counterparts. We quantify the influence of activity on aggregates topology, as they transition from branched structures to unconnected chains and rings. Our findings are summarized in a state diagram, delineating the impact of dipolar coupling strength and active force magnitude on the system.
\end{abstract}
\maketitle

\section{\label{sec:level1}Introduction} 
Active matter is an interesting class of non-equilibrium materials, in which  every individual constituent  is  capable of  energy uptake from the environment  and   converting it to some kind of directed movement or work~\cite{ramaswamy_mechanics_2010, marchetti_hydrodynamics_2013}. Active systems span  a wide range of length scales from microscopic  subcellular bio-filaments propelled by molecular motors and bacteria to macroscopic fire ant rafts and bird flocks. 

Active particles interact  via direct, {\it e.g.}  steric or contact patchy interactions~\cite{active_patchy},  and long-range and often anisoropic field-mediated forces such as hydrodynamic~\cite{marchetti_hydrodynamics_2013,elgeti_2015}, chemotactic~\cite{stark1,popescu2018chemotaxis,Liebchen_19,Fadda_2023} or electromagnetic fields~\cite{yan2016reconfiguring,Quinckie_rollers,ren2012assembly,magbot_21}. As a result, they exhibit rich patterns of  collective behavior~\cite{collective_motion_Vicsek} such as dynamic clustering, flocking~\cite{Vicsek} and motility-induced phase separation~\cite{theurkauff,palacci,MIPS}.

  The increasing interest in collective properties of  such intrinsically out of equilibrium self-organized structures has fostered the engineering of synthetic active particles whose self-propulsion and interactions can be controlled. Active colloids~\cite{activ_colloid_bishop_23}, nano- and micro-robots~\cite{magbot_21} are prominent examples of man-made self-driven particles that generate  self-propulsion through different motives \cite{activ_colloid_aranson2013,activ_colloid_ebbens2016,activ_colloid_Zhang_2017,activ_colloid_ebbens2016}, {\it e.g.} via electric, magnetic fields~\cite{Quinckie_rollers,ren2012assembly}, light~\cite{palacci}  or chemical gradients~\cite{theurkauff}. The possibility  to control  motility of man-made active particles as well their symmetry and range of interactions makes them promising candidates  for   the development of innovative applications in diverse fields, ranging from materials science to  creation of artificial nano- and micro-machines for performing functional tasks.

In the pursuit of generating novel active materials, it is crucial to comprehend consequences of interplay between activity and interparticle interactions on their emergent collective properties. While some consensus has been reached regarding the impacts of short-range interactions, such as excluded volume and contact attractive interactions, on non-equilibrium structure formation~\cite{abp_Romanczuk2012,hagan_2013,Critical_ABP,MIPS_Caprini_20,Caporusso2020motility,Maggi_21,Caprini_22,saric,mallory2018active,attractive_colloids}, the effects of long-range interactions on the dynamic self-assembly of active particles, especially in 3D, remains considerably less elucidated.
An important example of long-range anisotropic interactions is dipole–dipole interactions arising from intrinsic dipole moments in  magnetotactic bacteria~\cite{magnetotactic_bacteria_Bazylinski,MTB_24} or in magnetic robots~\cite{magbot_21,magnetic_robotic_21} or induced dipole moments  generated by electric or magnetic fields in active colloids~\cite{ren2012assembly,yan2016reconfiguring,sakai2020active_dipolar_colloids}.

A notable feature of dipolar interactions is their dependence on both   orientations and spatial configurations of interacting particles. Specifically, when two particles are in a head-to-tail contact configuration, they display an absolute minimum in the two-body potential energy, whereas those arranged side-by-side with antiparallel alignment give rise to a relative minimum in dipolar energy, see Fig.~\ref{fig:scheme}.  Consequently, even in passive colloids,   long-range anisotropic dipolar interactions give rise to intricate patterns of structure formation~\cite{klapp_2016}.  At   low densities, dipolar particles self-assemble into rings, linear chains and a gel-like network of branched aggregates upon increase of dipolar coupling strength~\cite{mc_weis,Weis_94,Sciortino_2012} and at higher densities  they develop nematic and polar order~\cite{Order_weis,Weis_2005,Weis_dipol_order_2006}.

For active particles, one expects that the intricate nature of long-range dipolar interactions 
  to give rise to dynamic self-assembly behaviors markedly distinct from active polar particles with short-ranged  alignment interactions such  as in the Vicsek model~\cite{Vicsek}. Indeed, studies of dipolar active Brownian particles in 2D exhibit a rich dynamical self-assembly behavior~\cite{dipolar_active_abp_2d_klapp,Klumpp_2020}.
However, the dynamical self-assembly of active particles with dipolar interactions in 3D has received much less attention~\cite{sakai2020active_dipolar_colloids,traveling_string}. Here, we focus on Brownian dynamics simulations of low density active dipolar particles and we provide an in-depth characterization of their structural and dynamical features.

In the passive limit, our system interacting with truncated Lennard-Jones potential behaves like dipolar hard spheres~\cite{Sciortino_2012}. It exhibits chain-like structures and branched networks which percolate through the systems with increasing dipolar coupling strength. 
Activating particles with low to moderate forces, the  structure and topology of self-assembled structures  are maintained, but they are molten beyond a critical active force which depends on the dipolar coupling strength. Interestingly, at low active forces and sufficiently strong dipolar coupling strengths, we obtain active polymer-like and active gel structures, which exhibit  enhanced dynamics compared to their passive counterparts.  While active string fluids and gels exhibit  structures akin to their passive counterparts, the activity-induced decrease in the mean lifetime of dipole-dipole bonds accommodates more frequent configurational rearrangements, leading to   enhancement of both translational and rotational diffusion of particles in aggregated structures.

 The remainder of the article is organized as follows.  In section \ref{sec:method}, we present our model of dipolar active particles and details of simulations and analysis of data. 
We begin presenting our results in section \ref{sec:stat_diagram}, where we provide an overview of the observed steady-states summarized in a state diagram. Then, we discuss the structural and dynamical features of each state in sections \ref{sec:str}  and \ref{sec:dyn}, respectively. We conclude the paper with final remarks and future directions in section \ref{sec:conclusion}.
  

\section{Simulations and analysis  details} \label{sec:method}
\subsection{Model system and dynamical equations}
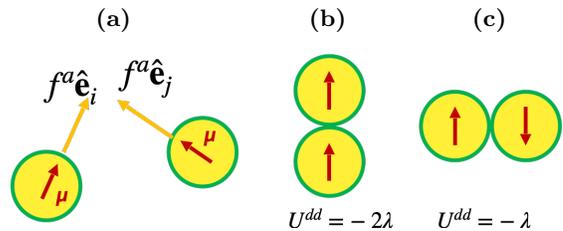
\begin{figure}
    \centering
\begin{tikzpicture}
            \draw (0,0) node[inner sep=0]{\includegraphics[width=0.85\linewidth]{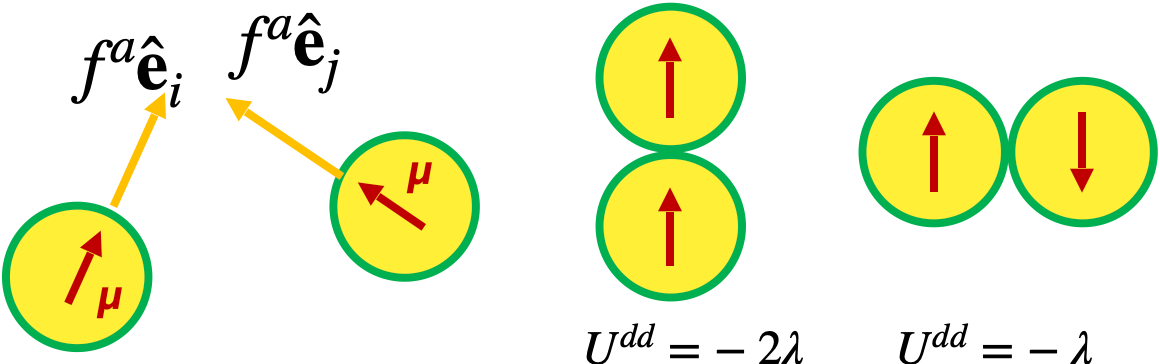}};
            \draw (-2.3,1.6) node {\textbf{(a)}};
            \draw (0.55,1.6) node {\textbf{(b)}};
            \draw (2.7,1.6) node {\textbf{(c)}};
    \end{tikzpicture}
    \caption{Schematics showing (a)  two interacting spherical active magnetic particles of diameter $\sigma$ where their orientation of  each particle  is assumed to be in the direction of its dipole moment $\vec{\mu}$; (b) and (c) two dipoles with in head-tail and anti-parallel contact configurations corresponding to the absolute and relative minima of dipole-dipole interaction energy. Here, $\lambda=\frac{\mu^2}{k_BT \sigma^3}$ defines the dimensionless dipolar coupling strength.}
    \label{fig:scheme}
\end{figure}
 We consider a system of $N$ active Brownian spheres of diameter $\sigma$, where each particle indexed by $i$ is endowed with a point  magnetic dipole moment $\vec{\mu}_i=\mu \vec{e}_i $  of arbitrary orientation $\vec{e}_i$ at its center. The self-propulsion is modeled, as in the active Brownian particle model~\cite{TenHagen_2011,abp_Romanczuk2012,ABP_2016} by an active force $\vec{F}^a_i=F^{a} \vec{e}_i$, which has a constant magnitude and is directed along the magnetic dipole moment $\vec{\mu}_i$, see Fig.~\ref{fig:scheme}(a). The dynamics of active force is governed by orientational fluctuations whose strength is set by the rotational diffusion coefficient $D_r$.  The pair potential between particles located at positions $\vec{r}_i$ and $\vec{r}_j$, and orientations $\vec{\mu}_i$ and $\vec{\mu}_j$ includes contributions from steric effects modeled by the isotropic Weeks-Chandler-Anderson (WCA) potential~\cite{WCA}  and orientation-dependent dipole-dipole interactions $U_{\text{dd}}$:
 \begin{equation}
    U_{ij}(\vec{r}_i,\vec{r}_j,\vec{\mu}_i,\vec{\mu}_j)= U_{\text{WCA}}(r_{ij})+U_{\text{dd}}(\vec{r}_{ij}\vec{\mu}_i,\vec{\mu}_j),
 \end{equation}
 where $\vec{r}_{ij}=\vec{r}_{j}-\vec{r}_{i}$.
 The WCA potential only depends on the distance between particles $r_{ij}$ and is given by
 \begin{equation}
    U_{\text{WCA}}(r)=4  \epsilon \left[  (\frac{\sigma}{r})^{12} -(\frac{\sigma}{r})^6+\frac{1}{4}\right], \:\:   r< r_{c },
\end{equation}
 where $\epsilon$ denotes the strength of pairwise interaction energy, and $r_{c} = 2^{1/6} \sigma$ is the cutoff distance such that $U_{\text{WCA}}(r > r_{cut}) = 0$.
 The dipolar interactions are given by
 \begin{equation} \label{eq:dipole}
    U^{dd}_{ij}=\mu^2\frac{\vec{e}_i \cdot \vec{e}_j-3 (\vec{e}_i \cdot \hat{\vec{r}}_{ij})(\vec{e}_j \cdot \hat{\vec{r}}_{ij})}{r_{ij}^3}
\end{equation}
The dipole-dipole interaction potential is long-ranged and its strength scales with magnetic moment  as $\mu^2$. In addition to distance, it depends on the angle between two particle orientations, as well as the angle of the interparticle distance vector  $\vec{r}_{ij}$ with their orientations resulting in non-reciprocal interparticle torques. A head-tail configuration of two dipoles leads to an absolute minimum of interaction energy, whereas an anti-parallel contact configuration corresponds to a relative minimum of $U^{dd}$, see Fig.~\ref{fig:scheme} (b) and (c).

 In the overdamped limit,  the equations of motion for each particle's position and orientation are given by the coupled stochastic differential equations:
 \begin{eqnarray} \label{eq:BD}
    \mathbf{\dot{r}}_i=\frac{1}{\gamma_t}(F^a \mathbf{\hat{e}}_i+\vec{F}_i)+\sqrt{2D_t}\boldsymbol{\Lambda}^t_i \\
     \mathbf{\dot{\hat{e}}}_i=\frac{1}{\gamma_r} \boldsymbol{\tau}_i \times \mathbf{\hat{e}}_i+\sqrt{2D_r}\boldsymbol{\Lambda}^r_i\times\mathbf{\hat{e}}_i,
\end{eqnarray}
in which $\vec{F}_i=\sum_{j \neq i}\nabla_{\vec{r}_i}U_{ij}$ and  $\vec{\tau}_i=\sum_{j \neq i}\vec{e}_i \times \nabla_{\vec{e}_i}U_{ij}$.
Here, $D_t$ and $D_r$ represent translational and rotational diffusion constants, respectively, given by
\begin{equation} \label{eq5.3}
    D_t=\frac{k_BT}{\gamma_t} \text{ and } D_r=\frac{k_BT}{\gamma_r}
\end{equation}
where $\gamma_{t}$ and  $\gamma_{r}$ are the translational and rotational drag coefficients. $\boldsymbol{\Lambda}^t_i$ and $\boldsymbol{\Lambda}^r_i$
are translational and rotational white noises with zero mean and unit variance, \emph{viz.}, $\langle \boldsymbol{\Lambda}_{\text{tr},i}(t)\boldsymbol{\otimes}\boldsymbol{\Lambda}_{\text{tr},i}(t')\rangle= \textbf{1}\delta(t-t')$ and $\langle \boldsymbol{\Lambda}_{\text{rot},i}(t)\boldsymbol{\otimes} {\boldsymbol{\Lambda}_{\text{rot},i}}(t')\rangle= \textbf{1}\delta(t-t')$.

\subsection{Units and simulation parameters} \label{sec:parameters}
We choose $\sigma$ as  the length unit  and $\tau=1/D_r$ as the time unit  and $k_B T=1/\beta$ as the energy unit, with $k_B$ and  $T$ being the Boltzmann’s constant and   the temperature, respectively. In these units the dimensionless parameters include $t^*=t/\tau=t D_r$, $r^*=r/\sigma$, $\epsilon^*=\beta \epsilon$, $\mu^*=\sqrt{\beta \mu^2 \sigma^{-3}}$, $D_t^*=D_t/(D_r \sigma^2)=1/\gamma_t^*$, $f^a= \beta F^a \sigma$ and $\rho^*=\rho \sigma^3$.  Additionally, we define the dimensionless self-propulsion speed as $v^a=\frac{F^a}{ \sigma D_r \gamma_t}$ and the dimensionless dipolar coupling strength as $\lambda={\mu^*}^2$.  We note that the dimensionless self-propulsion speed $v^a$ is identical to the commonly used P\'eclet number in the active Brownian particles literature~\cite{MIPS,ABP_2016}.
In our simulations, we set $\epsilon^*=1$, $\gamma_t^*=3$ and run simulations at the fixed density of $\rho^*=0.02$. Defining the mean volume fraction as  $\Phi=\rho^* \pi/6$ gives rise to the effective packing fraction  $\Phi=0.010$. We vary the dipolar coupling strength and the active force in the range $ 1\le \lambda  \le 16$ and  $ 0\le f^a \le 200$.

Brownian dynamics equations of motion given by Eqs.~\eqref{eq:BD} are  implemented in the Hoomd-Blue molecular dynamics software package~\cite{hoomd} using a cubic box with periodic boundary conditions
and integrated with a discrete time-step $dt^*=2 \times 10^{-4}$. 
The number of particles in the majority of simulations is $N=8788$, unless otherwise stated. This leads to a large enough box size  for the low density considered  here, such that the maximum magnitude of the dipolar potential $2 \lambda/r^3$ at half-box length $L=76.02$ is smaller than $10^{-3}$ and the Ewald summation correction is negligible, as verified by running simulations for a larger system size of $N=19652$ corresponding to $L=99.42$. The data of the larger systems for $\lambda=16$ were used to calculate the structure factor.

To  obtain a steady-state efficiently, we first equilibrated the passive samples of different magnetic moments starting by $\mu^*=1$, where the initial configuration was that of randomly oriented particles located on a face-centered cubic lattice. The   equilibrated configuration of $\mu^*=1$  was  used as the starting point for equilibration of larger values of magnetic moments sequentially. To ensure that the system is equilibrated, we used the block averaging method~\cite{block_averaging}. 
The equilibrated passive samples  were then used as the initial configuration for active dipolar colloids with the same magnetic dipole moment, where we increase the motility gradually from $f^a=0$ to larger values up to a maximum of $f^a=200$ for $\lambda=16$. 
A typical  simulation run   consists of $t=100-1000$ $\tau$
for reaching a steady state, followed by a
production period of $100-10000$ $\tau$. 

\begin{figure*}[ht]
\centering
    \includegraphics[width=0.95\linewidth]{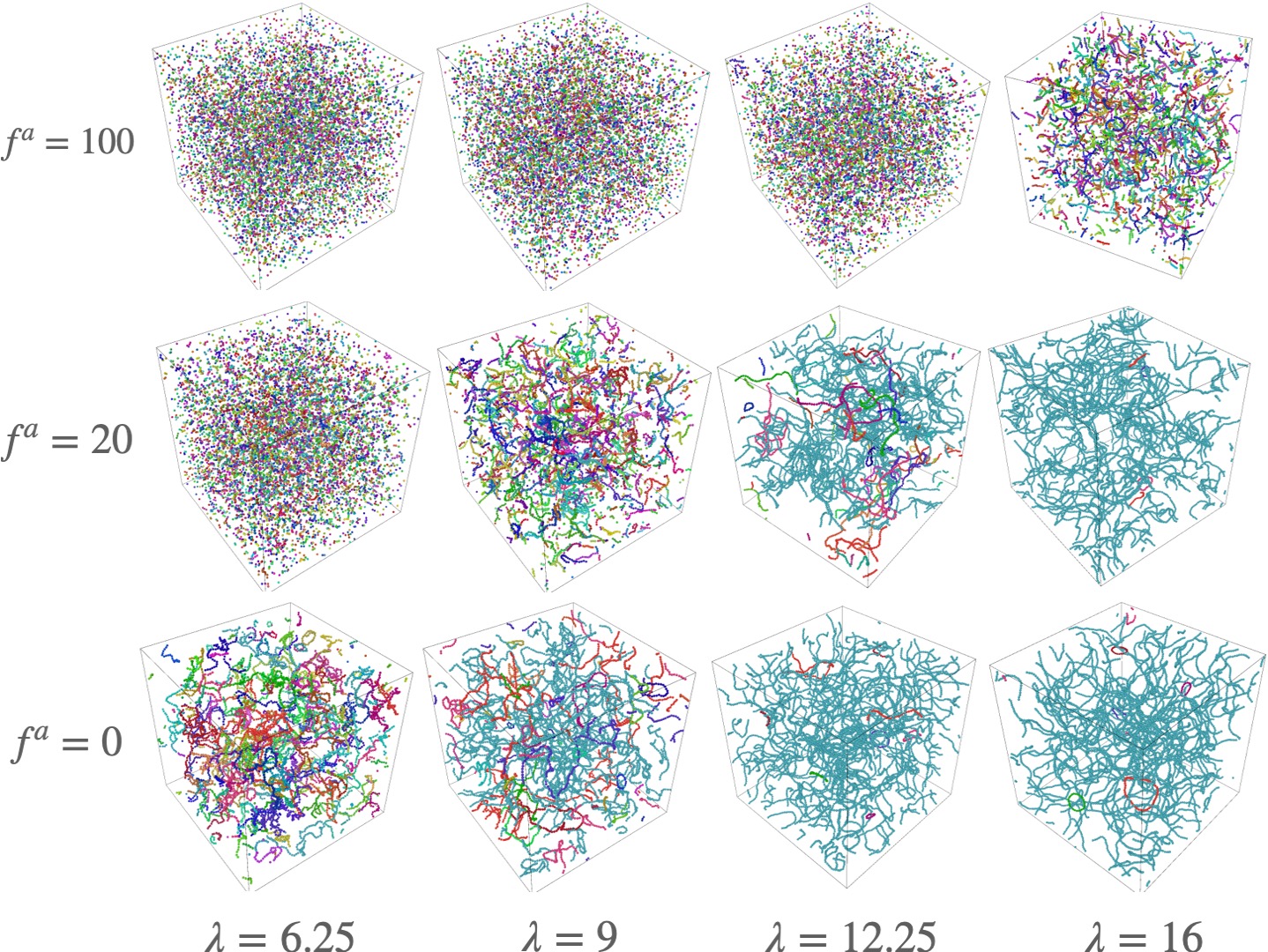}
     \caption{Representative snapshots of active magnetic particles interacting with varying dipolar coupling strengths at different activity levels, forming isotropic homogeneous gas-like fluid,  string fluid consisting of chain-like assemblies  and  percolated network structures. }
     \label{fig:snapshots}
 \end{figure*}

\subsection{Analysis of the configurations}
In this section, we provide details on our method of analysis of structural and dynamical features of collectives of active colloids and the definition of computed quantities.
 \subsubsection {Cluster analysis:  rings, chains and branched structures} \label{sec:cluster_analysis}
Passive dipolar spheres with sufficiently strong dipole-dipole coupling strengths have the tendency to form aggregates in the form of chains, rings and  branched networks. To quantify the affects of activity on the connectivity properties of the system, we perform a clustering analysis based on a simple distance criterion, similar to references~\cite{Fadda_2023,Sciortino_2012,dipolar_active_abp_2d_klapp}.  
 Two particles are considered as being neighbors
 if their center-to-center distance is smaller than a  threshold cutoff $r_t$. To determine the clusters in the system, we first identify the neighbors of each particle by using $k$-d trees~\cite{cormen2009Algorithms}. Then, we construct a unidirected graph that labels the connected particles by cluster-$id$ and cluster-$size$~\cite{SciPy,NetworkX}. This information is then used to distinguish    the ring, chain and branched topology of the clusters.
  Rings correspond to clusters where every particle has exactly two neighbors. In a chain, there exist exactly two particles with one neighbor and all the other particles have two neighbors.  The clusters with branched topology are those clusters in which at least one particle has more than two neighbors. We also identify a percolating network of aggregates as an interpenetrated cluster which connects at least two opposite sides of the simulation box.
  
Based on size distributions of different cluster topologies, we constructed several useful quantities. We first calculated the normalized  probability distribution   of clusters of any size $n$ with topology $\alpha$ corresponding to ring, chain or branched structures, as $P^{\alpha}(n)=N_c^{\alpha}(n)/\sum_{n=1} N_c^{\alpha}(n)$, where $N_c^{\alpha}(n)$ is the number of clusters of  size $n$ with the topology  $\alpha$. The topology $\alpha$  of clusters is denoted by  $\alpha=c$ for chains, $\alpha=r$ for rings and $\alpha=b$ for branched structures.
The mean size of clusters with topology $\alpha$ for any snapshot is given by 
\begin{equation}
  N_{\alpha} (t)= \sum_{n=n_{min}}n P^{\alpha}(n) 
\end{equation}
where $\sum_{n=n_{min}} P^{\alpha}(n)=1$ and $n_{min}=2$,3 and 4 for chains,  rings and  branched structures, respectively. We obtain the time-averaged cluster size of topology $\alpha$ in the production runs denoted by $\langle N_{\alpha} \rangle$. Moreover, from distribution functions  we compute  the fraction of particles in ring, chain and branched configurations denoted by $\phi_r$, $\phi_c$ and $\phi_b$. We note that $\phi_f+\phi_r+\phi_c+\phi_b=1$, where $\phi_f$ corresponds to the fraction of free unbounded particles in monomeric state. 

We  define  the fraction of polymerized material $\phi_p$ by counting the fraction of particles belonging to any cluster of size $n \ge 2$,  regardless of  its topology:
\begin{eqnarray}
\phi_{p}= \frac{\sum_{n >1}\sum_{\alpha}n N_c^{\alpha}(n)}{N}=1-\frac{N_{mon}}{N}
\label{eq:phi_p}
\end{eqnarray}
where $N_{mon}$ is the number of particles in monomeric form not belonging to any kind of cluster.
We also determine the mean number of particles in the largest cluster, irrespective of  its topology, $N_{\text{c}}^{\text{max}}$ by  averaging over many steady-state snapshots in the production runs. The mean fraction of particles in the largest cluster 
\begin{eqnarray}
\phi_{\text{max}}= \frac{\langle N_{\text{c}}^{\text{max}} \rangle}{N} 
\label{eq:Nmax}
\end{eqnarray}
can be used as an order parameter~\cite{MatozPerco,dipolar_active_abp_2d_klapp} to characterize the percolation transition of chain-like aggregates. $\phi_{\text{max}}$ approaches zero in a state where fluid  mainly consists of unbounded particles  whose positions are  essentially uncorrelated, as reflected by a weak pair correlation function. Furthermore, $\phi_{\text{max}}$ is close to zero in a string fluid state where the mean size of assemblies, in forms of chains, rings or branched structures, is much smaller than $N$. In contrast, $\phi_{\text{max}} \to 1$ when the particles self-assemble into a system-spanning network of branched structures, with the cluster size comparable to $N$. 
We define a  string fluid (SF) as a fluid where more than 50\% of particles are in polymerized form, {\it i.e.} $\phi_p > 0.5$ and  $\phi_{\text{max}}< 0.7$. In our simulations, a percolated network (PN) is identified when $\phi_{\text{max}}\ge 0.7$, which aligns with the visual observations of system-spanning branched structures.
We choose $r_t=1.2$ for all the samples with $\rho^*=0.02$ based on the position of  the first minimum of the radial distribution function.  However, we checked that the cluster size distribution is not significantly affected by the exact value of $r_t$.

\subsubsection{Dynamical features}

In order to investigate the bond dynamics, we compute a bond time autocorrelation function (TACF) by assigning a bond variable \(n_{ij}(t)\) at each time \(t\) for every pair of particles \(i\) and \(j\), as described in reference \cite{bondTACF}. This variable is set to 1 if particles \(i\) and \(j\) are neighbors according to the criterion introduced above, and 0 otherwise. The bond TACF is then calculated as

\begin{equation}
    C_b(t)=\frac{2}{N(N-1)} \sum_{i=1}^{N-1} \sum_{j=i+1}^N\langle n_{ij}(t'+t) \cdot n_{ij}(t') \rangle_{t^{'}}
    \label{eq:C_b}
\end{equation}
where $\langle \cdots \rangle_{t^{'}}$ indicates averaging over all possible starting times $t'$.
This method allows for a comprehensive analysis of the structures' persistence in time and is crucial for understanding the mechanical properties of such systems~\cite{gel_bond_life}.

To characterize the orientational dynamics of active particles, we compute the time autocorrelation function of their unit orientation vectors $\boldsymbol{\hat{e_i}}$
\begin{equation}
    C_e(t)=\frac{1}{N} \sum_{i=1}^N \langle \boldsymbol{\hat{e_i}}(t'+t) \cdot \boldsymbol{\hat{e_i}}(t) \rangle_{t^{'}}
    \label{eq:C_e}
\end{equation}
referred to as orientational TACF. In the infinite dilution limit where the orientations of  Brownian particles  diffuse independently, $C_e(t)$ decays as $\exp(-2D_rt)$.

To quantify the translational dynamics of active particles, we first evaluate the mean-squared displacement of particles  obtained as
\begin{equation}
  \langle \boldsymbol{\Delta} \vec{r}^2(t) \rangle=  \frac{1}{N}  \sum_{i=1}^N \langle [\vec{r}_i (t'+t)-\vec{r}_i (t')]^2 \rangle_{t'}. 
  \label{eq:MSD}
\end{equation}
 In addition, we compute the self intermediate scattering function  defined as 
\begin{eqnarray}
    f_s(\vec{q},t)&=&\frac{1}{N} \sum_{i=1}^N \left \langle  \exp[i\vec{q}\cdot (\vec{r}_i (t'+t)-\vec{r}_i (t'))] \right \rangle_{t'}.  \label{eq:F_s}
\end{eqnarray}
 $f_s(\vec{q},t)$ is also referred to as the incoherent scattering function and it describes the average single particle diffusive motion in the reciprocal space. 

\begin{figure}[h]
\centering
    \includegraphics[width=\linewidth]{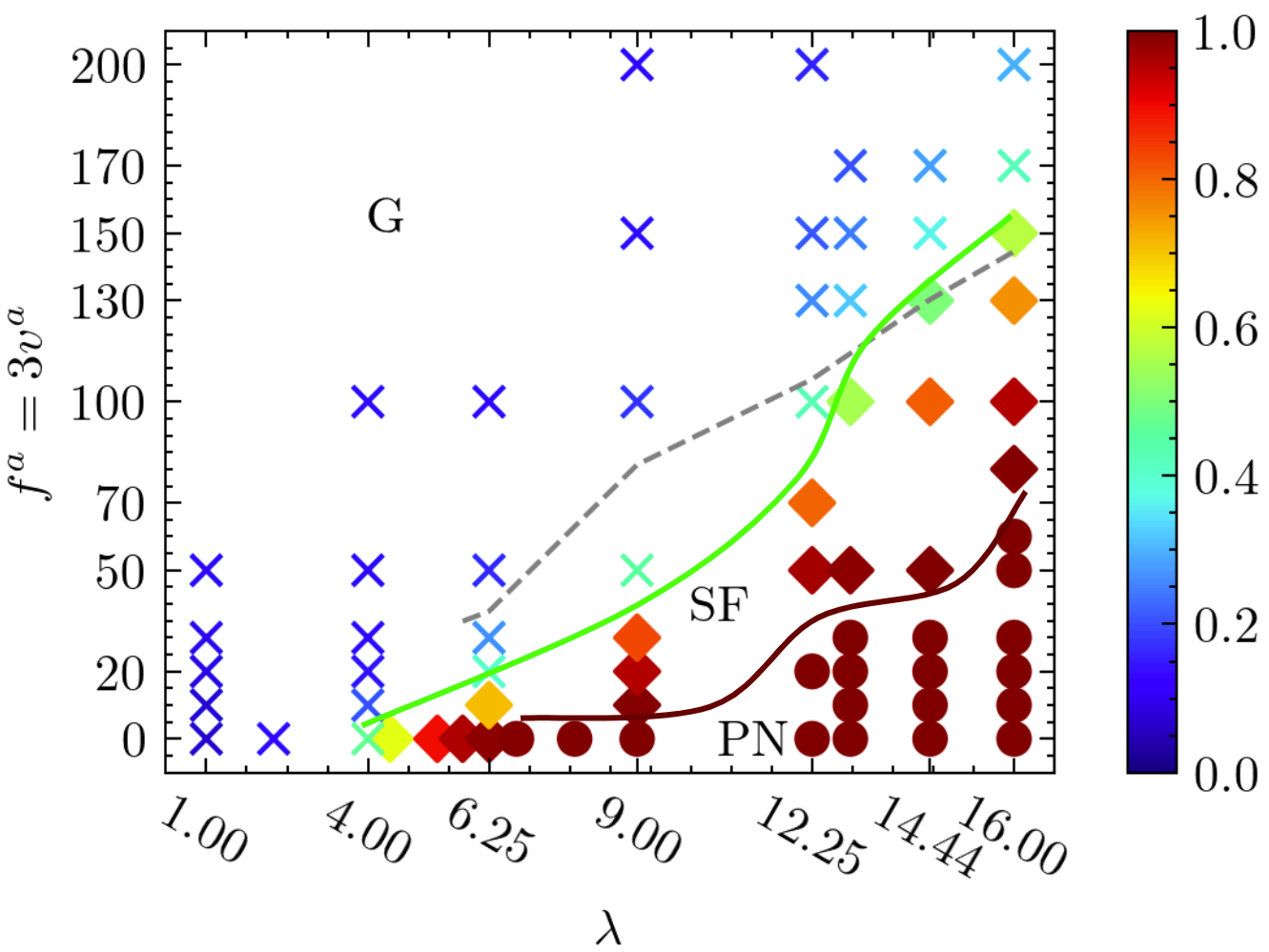}
    \caption{Non-equilibrium state diagram of active dipolar colloids in 3D as a function of coupling strength $\lambda$ and reduced active force $f^a$. We distinguish three different states,  a gas-like fluid (G) mainly consisting of  unbounded particles denoted by crosses $\times$, a string fluid  (SF) consisting  primarily  of chains and rings denoted by diamonds, and percolated network (PN) represented by disks. The color bar encodes the mean degree of polymerization $\phi_p$. The dark red line separates the PN states from SF states and the green line shows where aggregates are destroyed by large active forces.  
   }
     \label{fig:phase_diagram}
\end{figure}
 
\section{Non-equilibrium steady-state diagram} \label{sec:stat_diagram}
In this section, we provide an overview of  our  results from  BD simulations  for active particles with dipole-dipole interactions, while varying the dipolar coupling strength $0 \le \lambda \le 16$ and  dimensionless active force  $ 0\le f^a \le 200$ at a  fixed density $\rho^*=0.02$, corresponding to $\Phi=0.01$.
We start by  a pictorial representation of  steady-state configurations of active magnetic particles  as we vary the $\lambda$ and $f^a$ values  presented in Fig.~\ref{fig:snapshots}.  
For passive systems $f^a=0$, at $\lambda=6.25$ we observe mainly chain-like structures and a few rings. As we increase the dipolar coupling strength, more branched structures appear in the system and the number of chains decreases such that at the highest $\lambda$ the system primarily consists of a percolated branched network coexisting with a few rings and chains. The equilibrium structures   at $f^a=0$  are in agreement with the trend observed for dipolar hard spheres when decreasing the temperature~\cite{Sciortino_2012,mc_weis}. 
When activating the dipolar particles, a general pattern emerges: there is a  coupling strength-dependent active force above which self-assembly  is inhibited, leading the system to transition into an active gas state. 
 Hence, overall active systems have  smaller mean cluster sizes. The effect is specially notable at the lowest $\lambda=6.25$, where already at $f^a=20$, chain aggregates are predominantly destroyed. However, as we increase $\lambda$, a larger active force is required to overcome the dipolar attraction and break the chain-like structures, in agreement with findings of dipolar active Brownian particles in two dimensions~\cite{dipolar_active_abp_2d_klapp}. Particularly at the highest coupling strength $\lambda=16$, the system keeps its interconnected network structure at low active forces leading to formation of an  \emph{active gel}, whereas  at moderate activity levels \emph{active polymer-like} structures predominate the system. Eventually, at sufficiently high active forces the system transforms into an isotropic fluid.

Fig.~\ref{fig:phase_diagram} presents the state diagram as functions of active force $f^a$ and dipolar coupling strength $\lambda$.  Here, various non-equilibrium structures are delineated based on target  quantities obtained from the cluster analysis as detailed out in section \ref{sec:cluster_analysis}. In summary, we distinguish three distinct states, a gas-like isotropic fluid (G) characterized by $\phi_{p} <0.5$, an isotropic string fluid (SF) consisting  primarily of chain-like structures identified as $\phi_{p} > 0.5$ and $\phi_{\text{max}}< 0.7$, and a percolated network (PN) for which  $\phi_{p} \rightarrow 1$ and $\phi_{\text{max}} \geq 0.7$.
At low dipolar couplings  $\lambda <6$,   active
particles at such a low density typically display a homogeneous
isotropic fluid state consisting mainly of monomers, without any significant structure formation,  which we refer to as an active gas. For $\lambda > 6$, an active force exceeding the $\lambda$-dependent threshold $f^a_{t1}$ is required to overcome dipolar forces bonding the colloids resulting in an active gas, see the green line in Fig.~\ref{fig:phase_diagram}. As can be seen from  Fig.~\ref{fig:phase_diagram}, $f^a_{t1}$ rises with an increase of $\lambda$. Its value can be roughly estimated as the active force which can overcome the dipolar attractive force between two aligned dipoles in head-to-tail configuration  at their mean nearest neighbor distance  $r_{nn}$.
The $r_{nn}$ can be determined from the first peak of the radial distribution function, see Sec. \ref{sec:gr}. The magnitude of the dipolar attractive force at  $r_{nn}$  in head-to-tail configuration is given by $6 \lambda / r_{nn}^4 $ and is shown by a dashed line in Fig.~\ref{fig:phase_diagram} predicting roughly the active force $f^a_{t1}$ where system transitions to an active gas. 

For intermediate dipolar coupling strengths $6 < \lambda < 9$ and $ 0<f^a < f^a_{t1}(\lambda)$, dipolar particles form   chains and rings, constituting an active string fluid. For larger dipolar coupling strengths  with $\lambda > 9$  and active forces below a critical $\lambda$-dependent active force, $f^a < f^a_{t0}(\lambda)$, demarcated by the dark red line in Fig.~\ref{fig:phase_diagram},  the system stabilizes into an active gel consisting of a percolated network of branched chain structures.  For  intermediate active forces  in the range $f^a_{t0}(\lambda) < f^a < f^a_{t1}(\lambda)$, the percolated network transforms  into a string fluid. Beyond $f^a_{t1}(\lambda)$, the  active polymer-like structures are disrupted  and the system transforms into an active gas. Having provided an overview of dynamical steady states, we quantify the effects of activity on structural and dynamical features of aggregates in the following subsections and their connectivity properties in the subsequent sections.
\section{Structural properties} \label{sec:str}
\subsection{Pair correlation function and structure factor} \label{sec:gr}
\begin{figure}
    \centering
    \begin{tikzpicture}
            \draw (0,0) node[inner sep=0]{\includegraphics[width=0.9\linewidth]{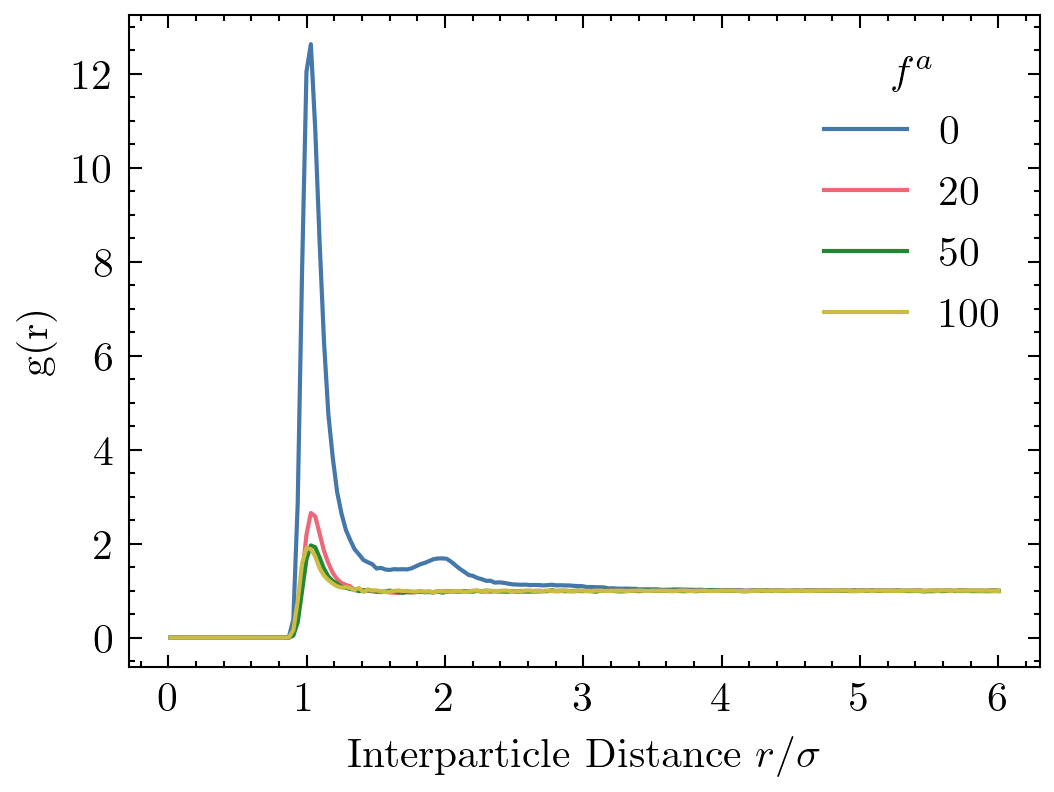}};
            \draw (-0.1,-6.) node {\includegraphics[width=0.9\linewidth]{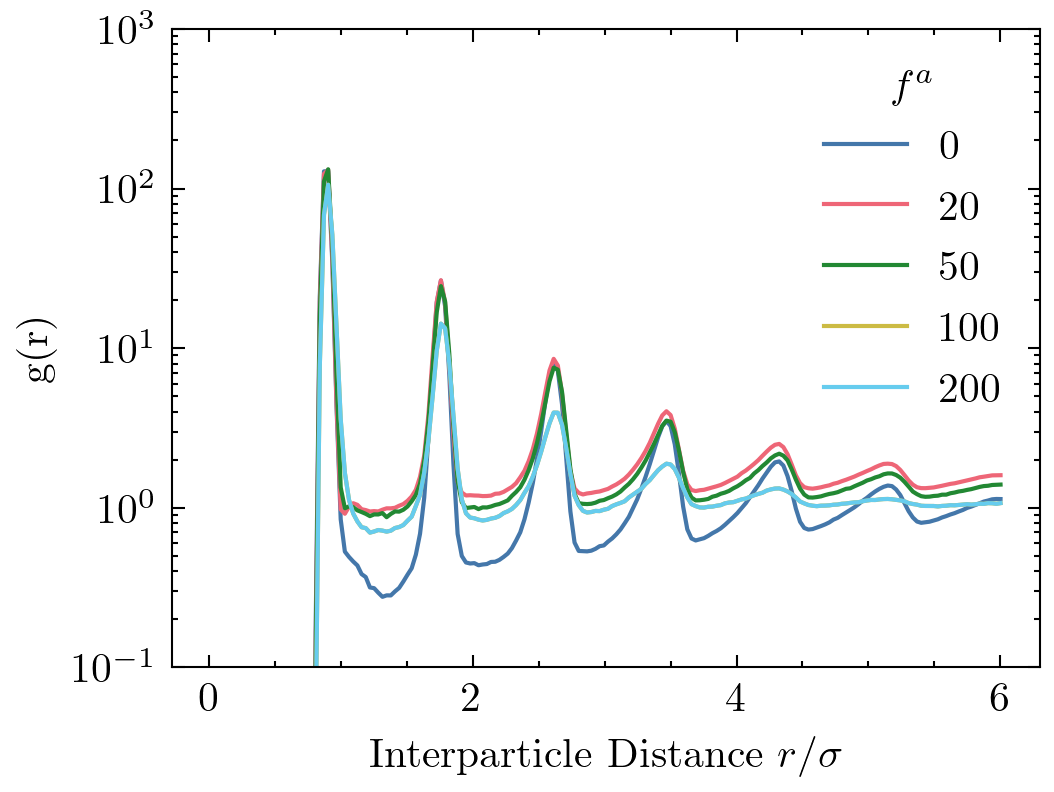}};
            \draw (-2.3,2.4) node {\textbf{(a)}};
            \draw (-2.3,-3.7) node {\textbf{(b)}};
    \end{tikzpicture}
    \caption{Radial pair distribution function for active dipolar particles at density $\rho^*=0.02$ at dipole-dipole coupling strengths a) $\lambda=4$ and b) $\lambda=16$ at different self-propulsion speeds $U_0^*=0$,20,50, 100 and 200.  }
    \label{fig:gr}
\end{figure}
We start by  quantifying the structural changes of the dipolar active particles at different activity levels. To this end,  we compute the radial pair distribution function $g(r)$, which provides information on the relative distance between particles in the system. Fig.~\ref{fig:gr} (a)  and (b) show   $g(r)$ at $\rho^*=0.02$ at two different dipole-dipole coupling strengths $\lambda=4$ and 16. At the lower coupling strength $\lambda=4$, in the passive limit the system mainly consists of small clusters with a mean size of about 2 which are primarily dimers. Hence, the pair correlation function shows a strong peak around $r/\sigma=1.0$ and a secondary peak around $r/\sigma=2$. Beyond this distance, $g(r)$ quickly approaches unity reflecting the  homogeneous isotropic nature of the fluid. Upon switching on the active force, the first peak becomes much weaker and the second one disappears as the systems transform into active gas-like fluids.
 At the higher coupling strength of $\lambda=16$, in the passive limit, $f^a=0$, the system consists of a system-spanning branched cluster which connects to itself via periodic boundary conditions, {\it i.e.} a percolated network. Hence, the $g(r)$ is well-structured and exhibits 5  peaks spaced regularly at $ 0.9n\sigma$ before approaching unity at large distances.  This shows that the system locally consists of straight chain segments up to about the 5th  nearest neighbors. For such a high coupling strength,  activating the particles does sustain  multiple peaks of $g(r)$   up to $f^a=100$. For $f^a \le 100$, heights and widths of the peaks change very little relative to those of passive particles, but the depths of minima of $g(r)$ decrease  with increasing $f^a$ due to active fluctuations.  For higher active force at $f^a=200$, the secondary peaks weaken and disappear beyond the 3rd neighbor, reflecting the breakage of chain-like structures at very large active forces and transition into an active gas.  
\begin{figure}
    \centering
    \begin{tikzpicture}
            \draw (0,0) node[inner sep=0]{\includegraphics[width=0.9\linewidth]{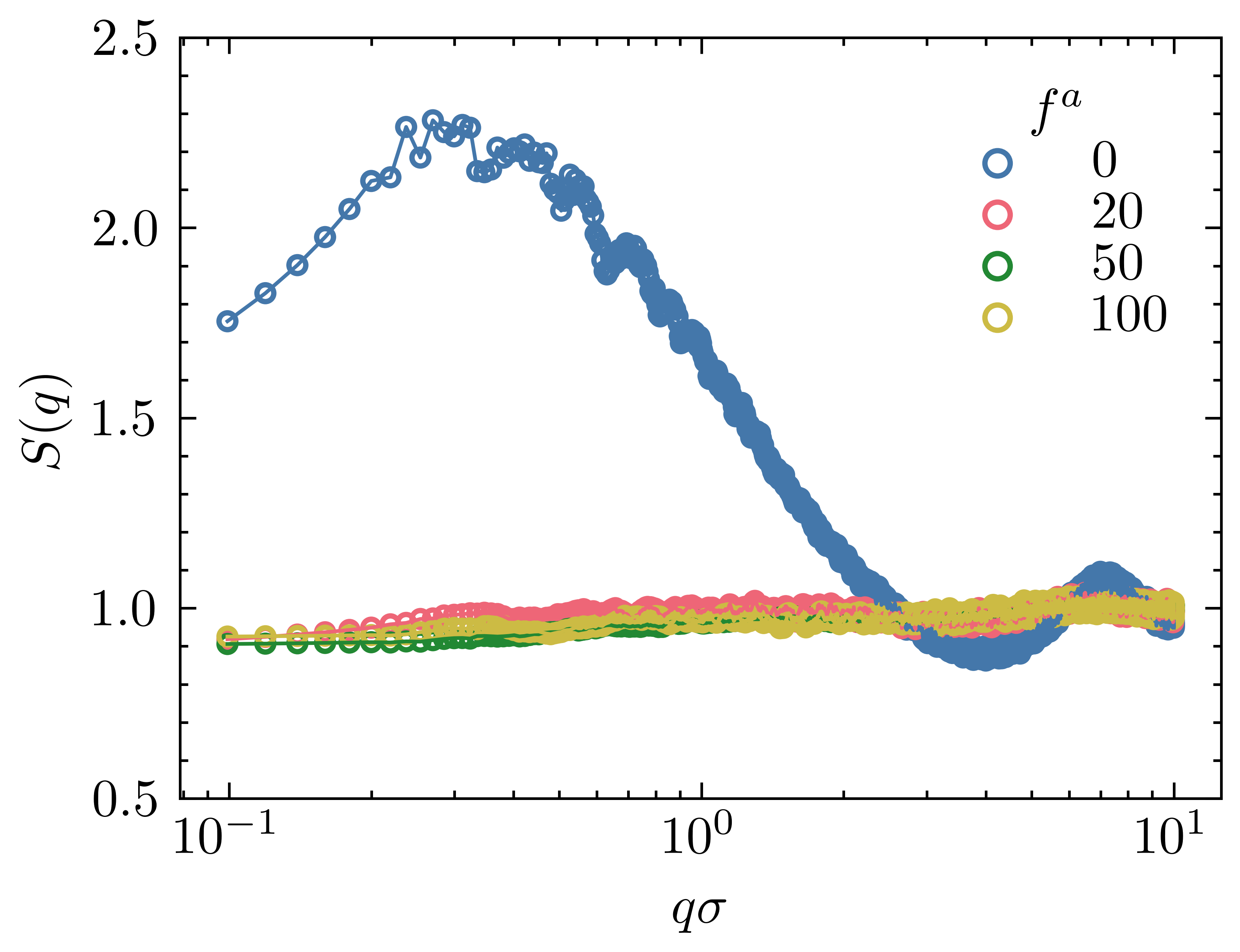}};
            \draw (-0.1,-6.) node {\includegraphics[width=0.9\linewidth]{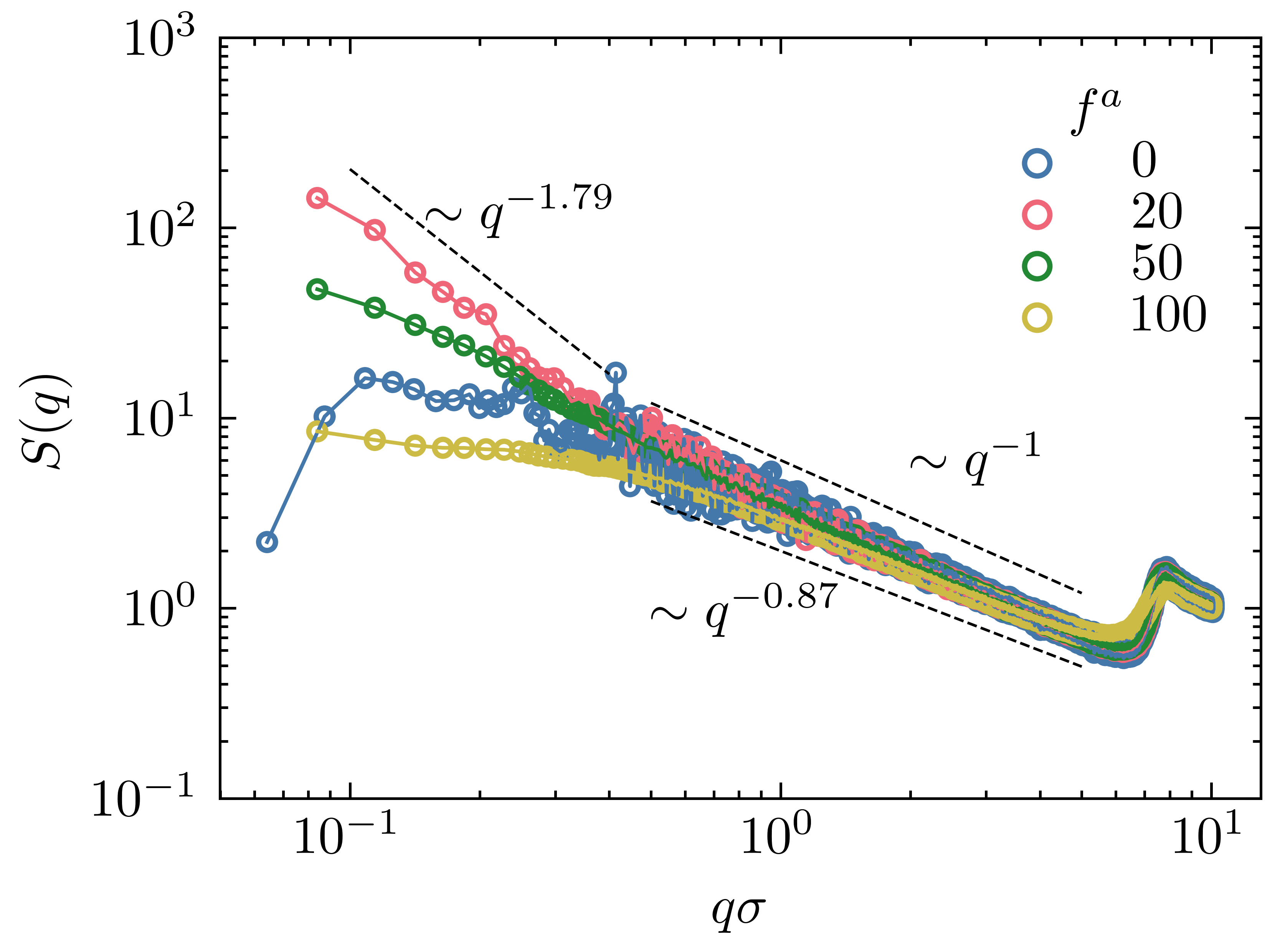}};
            \draw (-2.3,2.4) node {\textbf{(a)}};
            \draw (-2.3,-3.6) node {\textbf{(b)}};
    \end{tikzpicture}
    \caption{Structure factor of active dipolar particles at density $\rho^*=0.02$ at dipole-dipole coupling strength a) $\lambda=4$ and b) $\lambda=16$ at different self-propulsion speeds $U_0^*=0$,20,50, 100 and 200. }
    \label{fig:Sq}
\end{figure}
\\

Next, we compute the structure factor $S(q)$, defined as $\frac{1}{N}\langle \sum_{i,j}\exp(i \vec{q}\cdot \vec{r}_{ij})\rangle$ for $\lambda=4$ and 16  as presented in Fig.~\ref{fig:Sq} (a) and (b), respectively. For  $\lambda=4$ the passive  system is already an isotropic fluid, so activity has little effect on its structure factor. For the $\lambda=16$ case shown in Fig.~\ref{fig:Sq}(b)  in the passive limit $f^a=0$, we observe an asymmetric broad peak at   $q \approx 0.1$, reflecting the strong large-scale association of particles.
These peak heights seem  to be unaffected by the exact value of $\lambda$ once a percolated network is formed, see Fig.~\ref{fig:Sq_passive} in the appendix, indicating that the typical characteristic $S(q)$ of equilibrium gels is similar to the findings of reference~\cite{Sciortino_2012} for dipolar hard spheres. When increasing the active force, we observe remarkable changes in the shape of the structure factor at low $q$ values.  $S(q)$ at low wavenumbers for $f^a <100$ becomes elevated and for low $q$ values $S(q)$ exhibits a power law, reflecting  strongly correlated cluster structures for such a large coupling strength. 
In active gels, structural rearrangements induced by activity, such as bond breakage and formation, occur more frequently, see videos in the supplementary materials. Nonetheless, an active gel maintains its percolated network structure as evidenced by elevated $S(q)$ at small wavenumbers. 
 At intermediate $q$ values $ 0.5 < q\sigma < 5$, $S(q)$ does change very little  for  $f^a \le 50$ and it exhibits  a power law behavior   scaling as $S(q)\sim q^{- \alpha}$ where the exponent varies $ \alpha=0.97$. A slope close to unity at intermediate scales is a  signature of  a homogeneous network of linear aggregates~\cite{solomon_2010}  and  is consistent with the intermediate $q$ scaling found in prior studies of dipolar hard spheres~\cite{Sciortino_2012}.
However, at the high active force $f^a =100$, where the branched percolated network breaks down and only active self-assembled chains persist, the peak of $S(q)$ at low wave-vectors disappears and it becomes flatter for $q<0.5$. This trend indicates that there is less interference between different chain-like aggregates. However, for intermediate $q$ values ($0.5 < q\sigma < 5$), $S(q)$ still exhibits a power-law decay, albeit with a slightly reduced exponent of $\alpha=0.87$.

\subsection{Mean degree of polymerization and connectivity properties }
\begin{figure}[t]
    \centering
    \begin{tikzpicture}
            \draw (0,0) node[inner sep=0]{\includegraphics[width=0.9\linewidth]{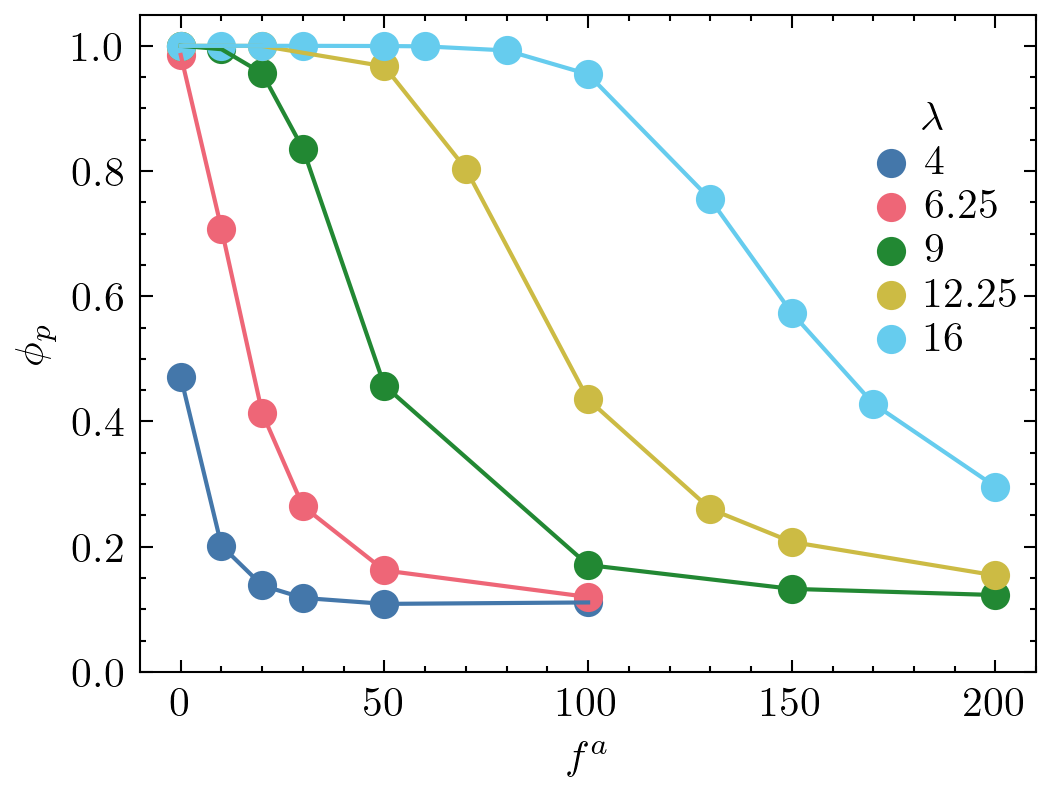}};
            \draw (-0.1,-6.) node {\includegraphics[width=0.9\linewidth]{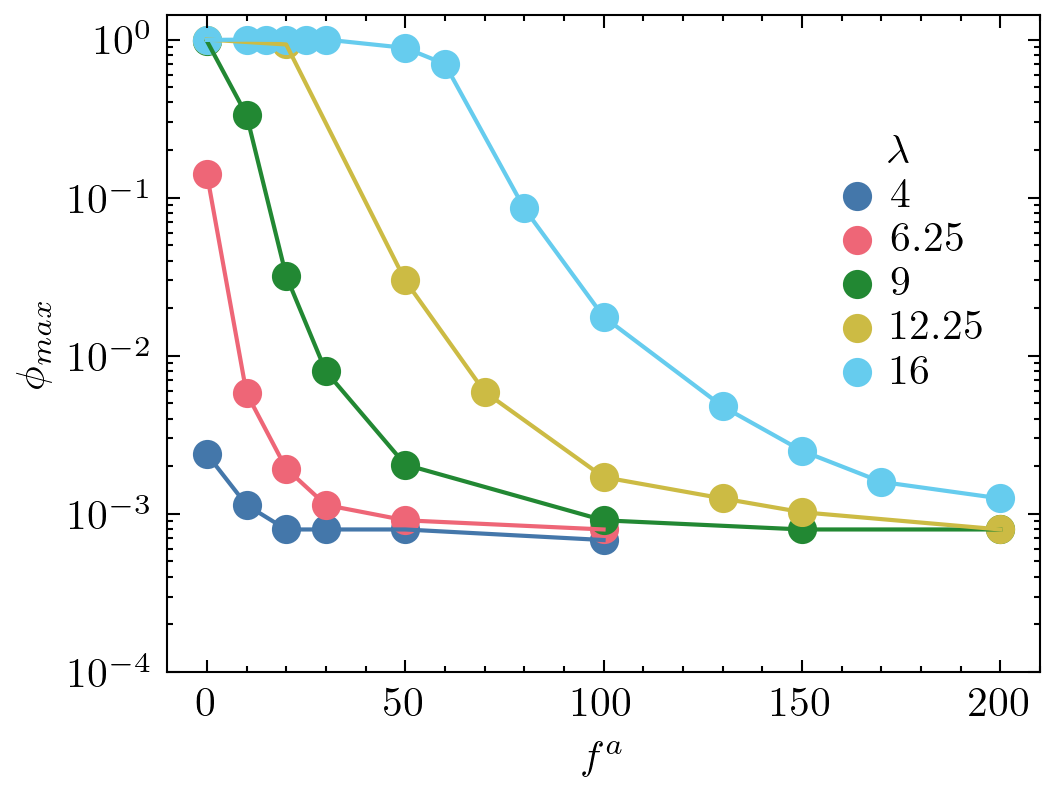}};
            \draw (3.4,2.6) node {\textbf{(a)}};
            \draw (3.4,-3.7) node {\textbf{(b)}};
            
    \end{tikzpicture}
\caption{a) The mean degree of polymerisation b) Mean fraction of particles in the largest cluster
as a function of reduced active force $f^a$  at different dipole-dipole coupling strengths $\lambda$.}
    \label{fig:assembly}
\end{figure}

To quantitatively characterize the effects of activity on self-assembly  behavior, we plot in Fig.~\ref{fig:assembly}(a) the mean degree of polymerization $\phi_p$, defined by Eq.~\ref{eq:phi_p}, 
as a function of the active force $f^a$ for  several coupling strengths in the range
$ 4 \le \lambda \le 16$. In the passive limit, at weak dipole–dipole interactions 
$\lambda=4$, the mean-degree of polymerization is about 0.5 and isotropic fluids mainly consist  of small clusters and monomers. However,  for coupling strengths   $\lambda > 6$, the majority of particles self-assemble and $\phi_p$ approaches unity. Increasing the activity level,  the  mean degree of polymerization decreases at a rate which depends on the strength of the dipole-dipole coupling $\lambda$. Especially, at   larger coupling strengths $\lambda > 12$, the system remains fully polymerized at low  active forces and we observe dissociation of assemblies only at  very large $f^a$  values.  

Next, we investigate the effect of activity on the mean fraction of particles in the largest cluster $\phi_{max}$, as defined in Eq.~\eqref{eq:Nmax}.   Fig.~\ref{fig:assembly}(b) presents $\phi_{max}$ versus active force $f^a$ at different $\lambda$. In the passive limit,  for $\lambda \ge 9$, $\phi_{max}$ approaches unity, see Fig.~\ref{fig:phi_max_lambda} in the appendix, implying formation of a system-spanning percolated network, in agreement with our visual observation. Upon switching on activity,  $\phi_{max}$ also decreases at a rate,  which depends on the dipole-dipole coupling strength. However, what is remarkable is the stability of the percolated network at the largest  coupling strength studied, {\it i.e.} $\lambda=16$,  up to moderate active forces $f^a \le 50$. 
 Having provided the information about the effect of active force on overall properties of assemblies, in the following we focus on detailed analysis of  distribution of particles in ring and chains.
 \begin{figure*}[t]
    \centering
    \begin{tikzpicture}
            \draw (0,0) node[inner sep=0]{\includegraphics[width=\linewidth]{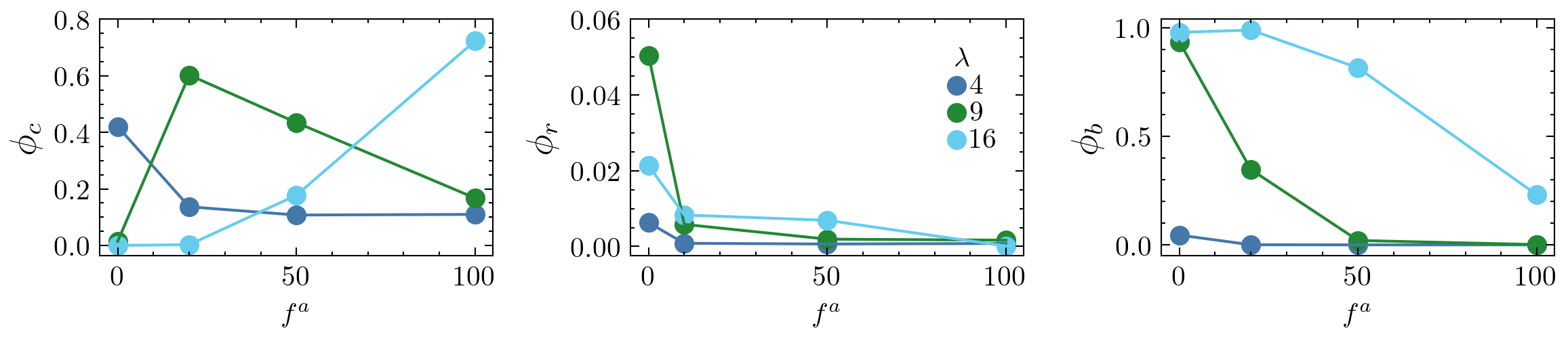}};
            
            \draw (-7.3,1.3) node {\textbf{(a)}};
            \draw (-1.,1.3) node {\textbf{(b)}};
            \draw (5.2,1.3) node {\textbf{(c)}};
            
    \end{tikzpicture}
    \caption{Mean fraction of particles in a) chain $\phi_c$, b) ring $\phi_r$  and c) branched structure $\phi_b$ configurations as functions of $f^a$  at dipolar coupling strengths $\lambda=4$,9 and 16. }
    \label{fig:phi_rs}
\end{figure*}
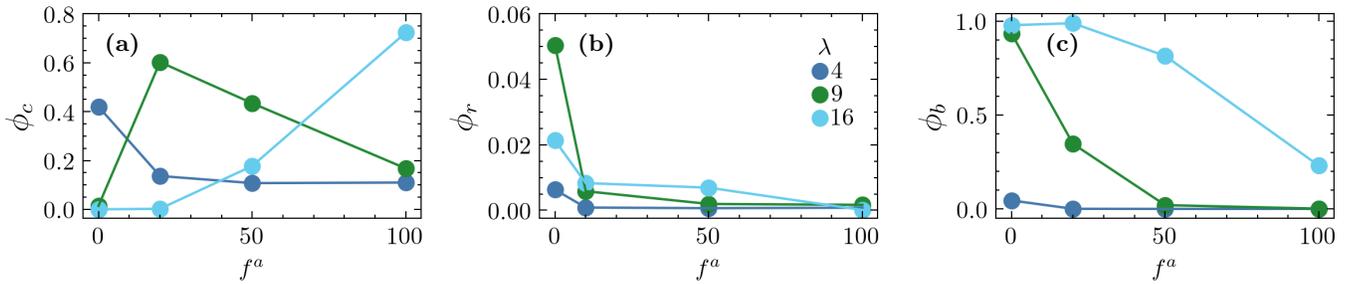

\graphicspath{{MOPIC/}}
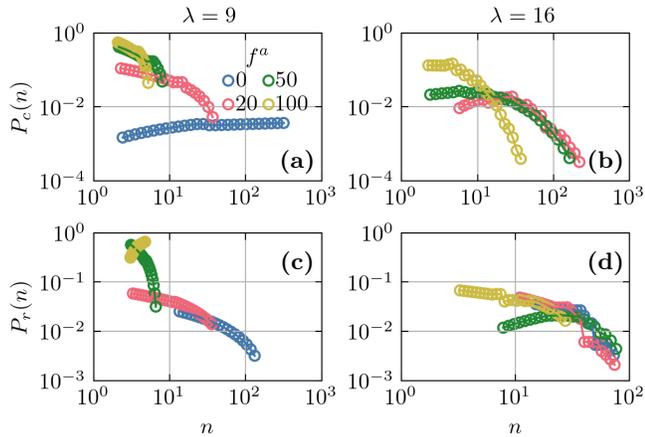
\begin{figure}[t]
    \centering
     \begin{tikzpicture}
            \draw (0,0) node[inner sep=0]{\includegraphics[width=\linewidth]{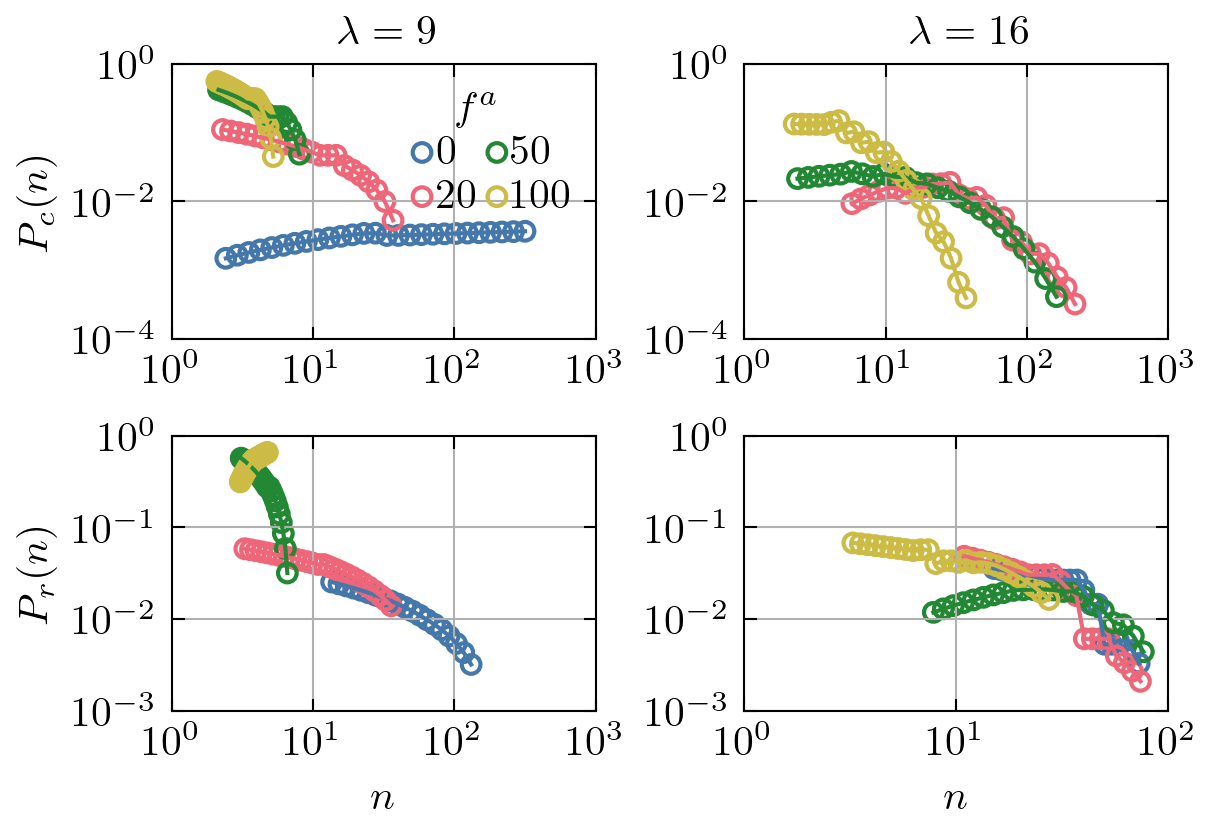}};
            \draw (-0.4,0.8) node {\textbf{(a)}};
            \draw (3.7,0.8) node {\textbf{(b)}};
            \draw (-.4,-0.55) node {\textbf{(c)}};
            \draw (3.7,-0.55) node {\textbf{(d)}};
    \end{tikzpicture}
    \caption{Probability distribution functions of chain length $n$ (a) and (b)  for $\lambda=9$ and 16
    and ring size $n$   (c) and (d)  for $\lambda=9$ and 16  for active forces $f^a=0$, 20, 50 and 100. } 
    \label{fig:P_Cring}
\end{figure}

 \begin{figure}[h!]
    \centering
    \begin{tikzpicture}
            \draw (-1.5,0) node[inner sep=0]{\includegraphics[width=\linewidth]{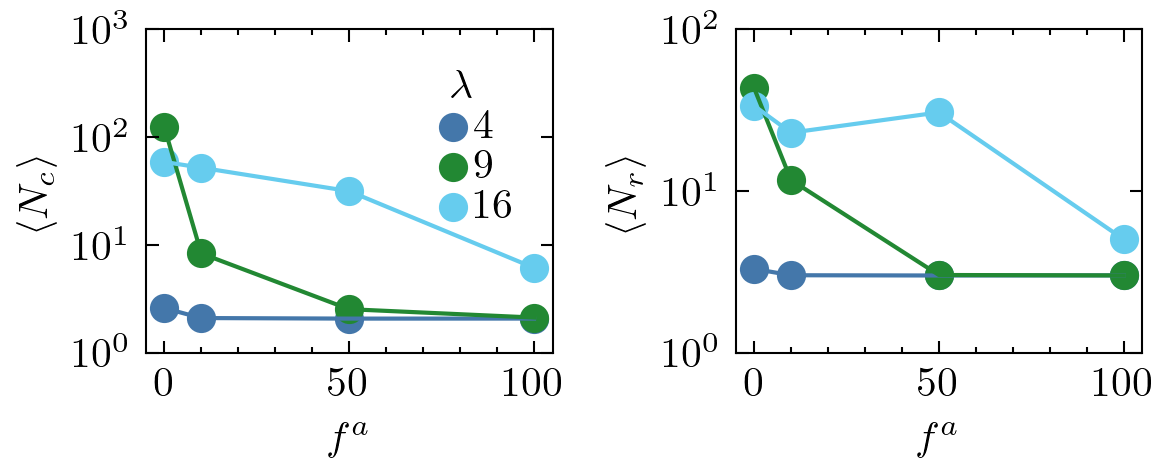}};
            
            \draw (-4.2,1.3) node {\textbf{(a)}};
            \draw (0.1,1.3) node {\textbf{(b)}};
            
    \end{tikzpicture}
    \caption{Mean size of a) chains $\langle N_c \rangle$ and b) rings $\langle N_r \rangle$ as functions of active force $f^a$ for different dipolar coupling strengths  $\lambda=4$,9 and 16.
    }
    \label{fig:mean-size}
\end{figure}

\subsection{Distribution of chains and rings}
We investigate the statistical properties of chains and rings  at different dipolar coupling strengths as we vary the activity level. 
We first investigate the mean fraction of particles in  chains $\phi_c$, rings $\phi_r$, and branched structures $\phi_b$ as functions of the active force $f^a$ at $\lambda=4$, 9 and 16, as presented in Fig.~\ref{fig:phi_rs}. 
At $\lambda=4$ in the passive limit, the system primarily consists of individual particles, with less than 1\% of particles in ring configuration, approximately 15\% in short chains, and about 5\% in small branched structures.
Upon increasing the active force, we find that fractions of particles in  chain, ring and branched configurations at $\lambda=4$  quickly drop down and an active gas of monomers is obtained. At $\lambda=9$ and 16, in the passive limit system forms a percolated branched network structure, resulting in  $\phi_b$ close to unity, which coexists with a small population of rings while $\phi_c=0$.

Upon increase of $f^a$, these systems evolve  from a percolated network structure to a string fluid and  finally an active gas, albeit at a slower pace for the higher coupling strength.
For $\lambda=9$ upon introduction of a small active force $f^a=20$, the system immediately transforms into an active string  (polymer-like) fluid as is visible by a notable increase of $\phi_c$, whereas the  fractions of particles in branched structures and ring configurations drop immediately. Upon further increase of  activity level,  fractions of particles in any from of self-assembled structure decrease as the system enters into an active gas state.
In the strong coupling regime, for $\lambda=16$  and sufficiently small active forces $f^a \le 50$, the branched network structure remains stable and even reinforced for small $f^a$, whereas the fraction of particles in chains increases slightly concomitant with decrease in the fraction of particles in rings. For  $f^a > 50$ the percolated network structure is destroyed and the transition to a string fluid state is visible by a notable increase of $\phi_c$ and an eminent decrease of $\phi_b$,  whereas the  fraction of particles in rings approaches zero.

Next, we investigate the probability distribution of  chain and ring sizes at $\lambda=9$ and 16, as plotted in Fig.~\ref{fig:P_Cring}. At these coupling strengths, collectives of dipolar active particles undergo a gradual transition, evolving from percolated network structures to active string fluids, and ultimately to active gas states as the activity level increases. In the passive limit, at $\lambda=9$ we have a few large strings and rings coexisting with a branched percolated network, whereas at $\lambda=16$ only rings coexist with a percolated network. Upon increase of active force, more chains and rings appear, but the distribution shifts toward smaller cluster sizes for both  $\lambda=9$ and 16, resulting in reduction of mean sizes of chains and rings as shown in Fig.~\ref{fig:mean-size}.

\section{Dynamical properties} \label{sec:dyn}
Having explored the effects of activity on  structural features of active magnetic particles, next we focus on its impacts on their dynamical properties.

\begin{figure}
    \centering
\begin{tikzpicture}
            \draw (0,0) node[inner sep=0]{\includegraphics[width=0.9\linewidth]{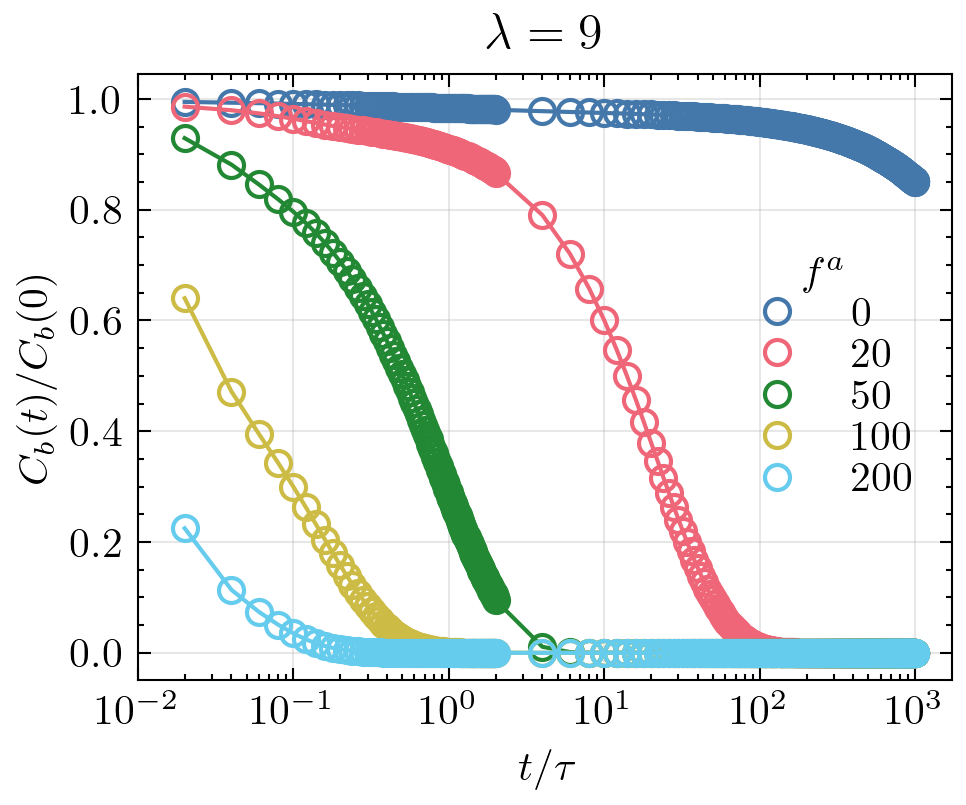}};
            \draw (0,-6.7) node {\includegraphics[width=0.9\linewidth]{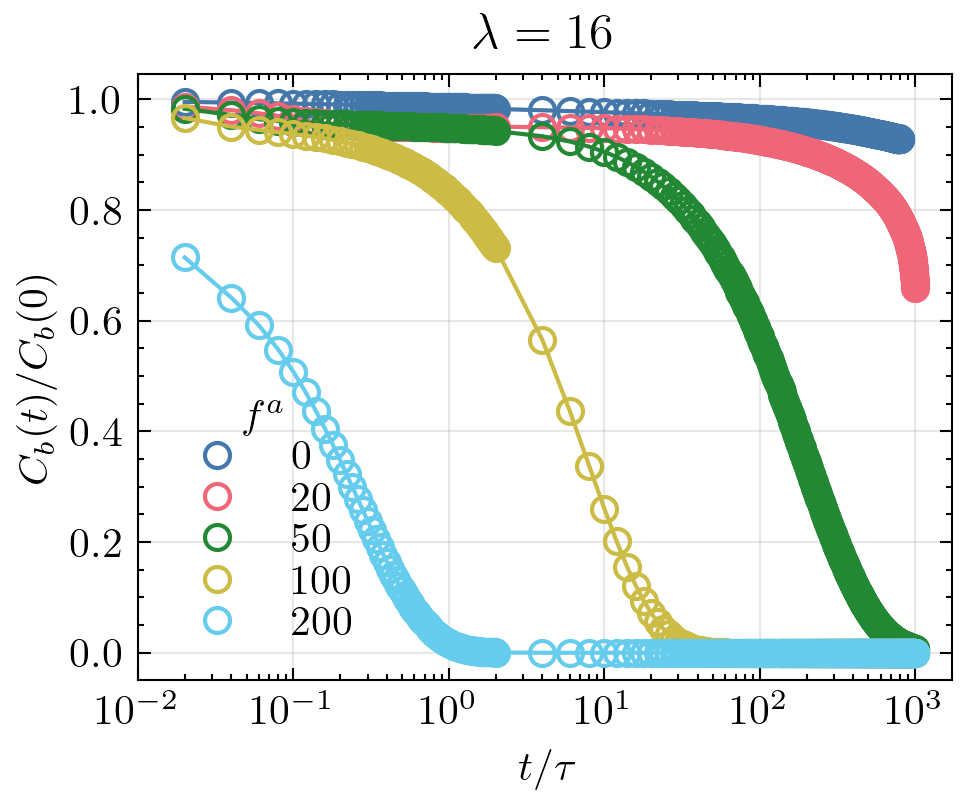}};
            \draw (-2.3,1.7) node {\textbf{(a)}};
            \draw (-2.3,-4.8) node {\textbf{(b)}};
    \end{tikzpicture}
    \caption{Bond time autocorrelation function $C_b(t)$ of active dipolar Brownian particles for coupling strength (a) $\lambda = 9$, and (b) $\lambda = 16$, at active forces $f^{a} = $ 0, 20, 50, 100, and 200.  
    }
    \label{fig:BCF}
\end{figure}
\subsection{Persistence of bonded structures}

We start by investigating the influence of activity on the dynamics of bonds formed due to dipole-dipole interactions.  Fig.~\ref{fig:BCF} shows the TACF of bond vectors, as defined in Eq.~\ref{eq:C_b}, normalized by its value at $t=0$ for dipolar coupling strengths $\lambda = 9$ and $\lambda = 16$, where the passive systems form percolated network structures at different active forces $f^{a}=$ 0, 20, 50, 100, and 200. At both coupling strengths,  for $f^{a}=0$ the presence of  percolated network is characterized by a rather slow-decaying $C_b(t)$, indicating very persistent bonds. Upon increase of active force, the $C_b(t)$s progressively decay   faster in both cases. Particularly at $\lambda = 9$, where we observe an immediate structural change from a percolated network to an active string (polymer) fluid already at $f^a=20$, this is well signaled by a notable decrease in the decay time of bond TACF. In the strong coupling regime with $\lambda = 16$, for low active forces $0 < f^{a} < 50$, where the system maintains a percolated network structure, the decay of $C_b(t)$ is observed to be faster compared to the passive case. However, the decay time still remains relatively large, indicating a more dynamic and time-reconfiguring network structure for active gels.
\begin{figure}
    \centering
    \includegraphics[width=0.9\linewidth]{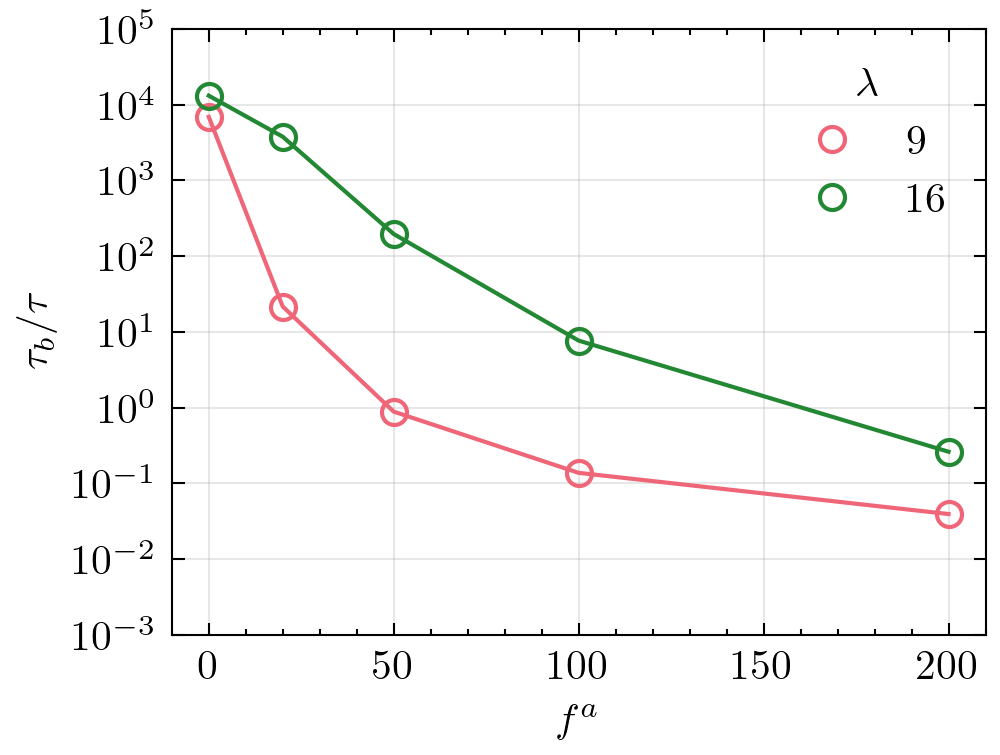}
    \caption{Characteristic decay time $\tau_b$ as a function of the active force $f^{a}$, obtained from the exponential fits $\exp(-t/\tau_b)$ of the bond time autocorrelation functions $C_b(t)$ for active dipolar Brownian particles at $\lambda = 9$ and $\lambda = 16$.}
    \label{fig:BCF_fit}
\end{figure}

  We find that all the $C_b$ curves can be well-described by an exponential decay law  of the form $\exp(-t/\tau_b)$. The $\tau_b$ parameter can be interpreted as the mean bond lifetime od dipolar particles. The evolution of $\tau_b$ with activity at both dipolar coupling strengths is presented in Fig.~\ref{fig:BCF_fit}. The mean bond lifetime decreases with $f^a$ for both cases and is  systematically larger for $\lambda = 16$ compared to $\lambda = 9$ at the same activity level.   The observed trends are  consistent with the structural transitions observed upon the increase of active force. Especially, when $\tau_b$ becomes of the order of unity, the systems transform into active gas states.

\begin{figure}
    \centering
\begin{tikzpicture}
            \draw (0,0) node[inner sep=0]{\includegraphics[width=0.9\linewidth]{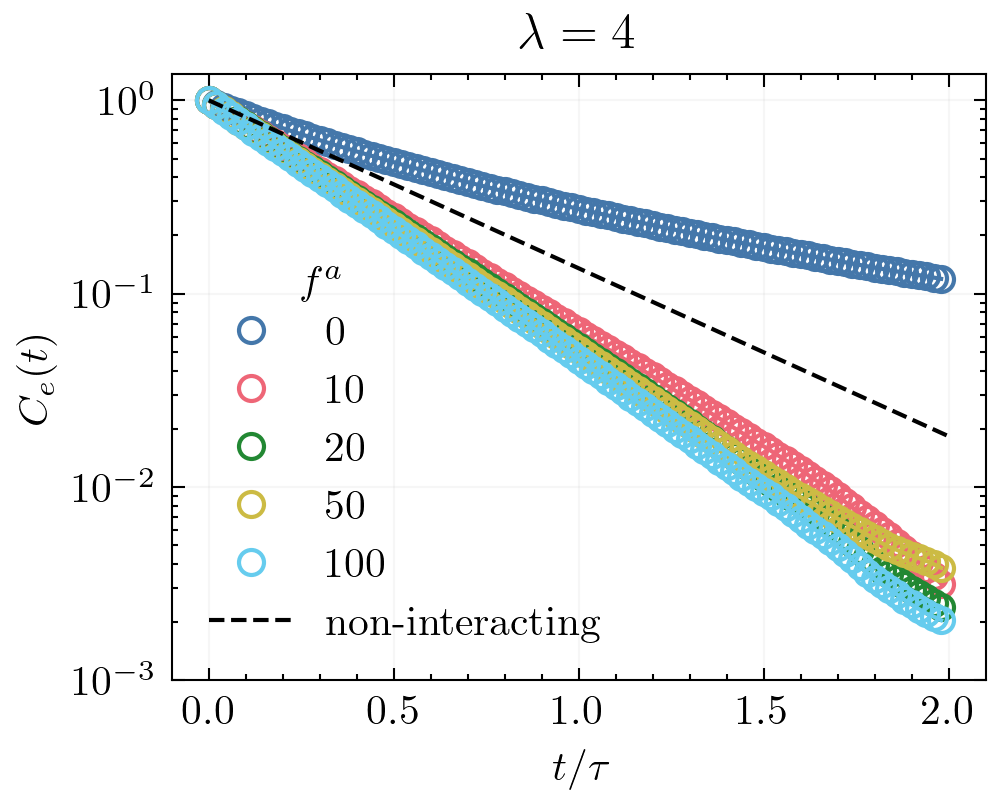}};
            \draw (0,-6.5) node {\includegraphics[width=0.9\linewidth]{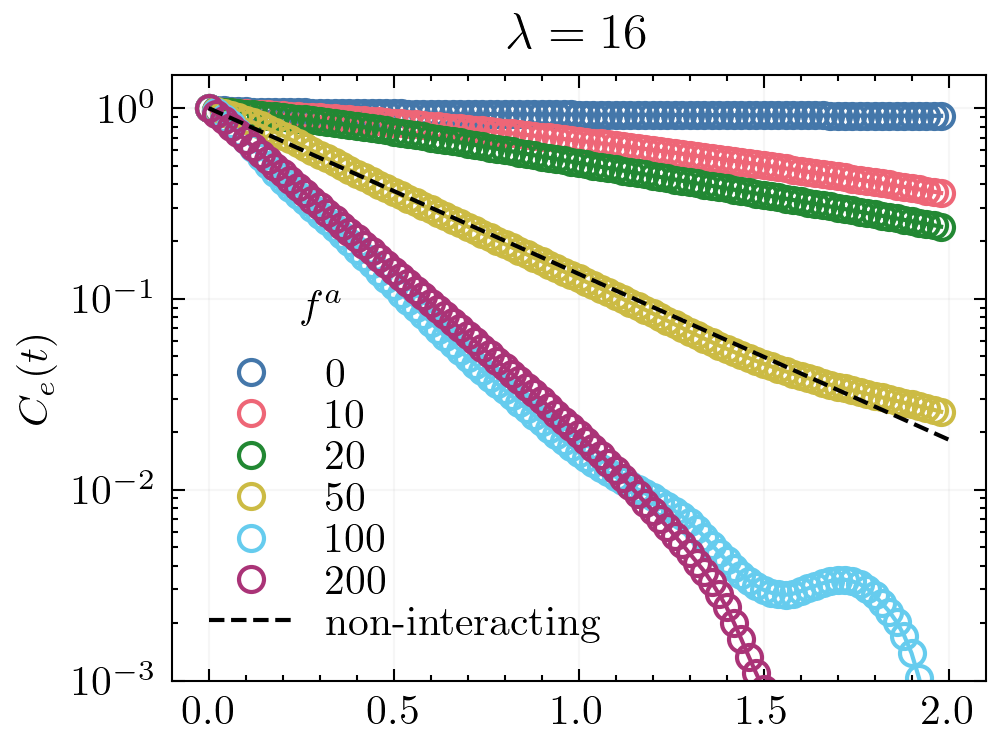}};
            \draw (-2.2,1.7) node {\textbf{(a)}};
            \draw (-2.2,-5.1) node {\textbf{(b)}};
    \end{tikzpicture}

    \caption{Orientational time autocorrelation function $C_e(t)$ of active dipolar Brownian particles for coupling strength (a) $\lambda=4$, and (b) $\lambda=16$, at active forces $f^a=$ 0, 20, 50, 100 and 200. Dotted lines correspond to the theoretical orientational TACF of a non-interacting, \textit{i.e} freely moving, Brownian particle.
     }
    \label{fig:OCF}
\end{figure}

\begin{figure}
    \centering
    \includegraphics[width=0.9\linewidth]{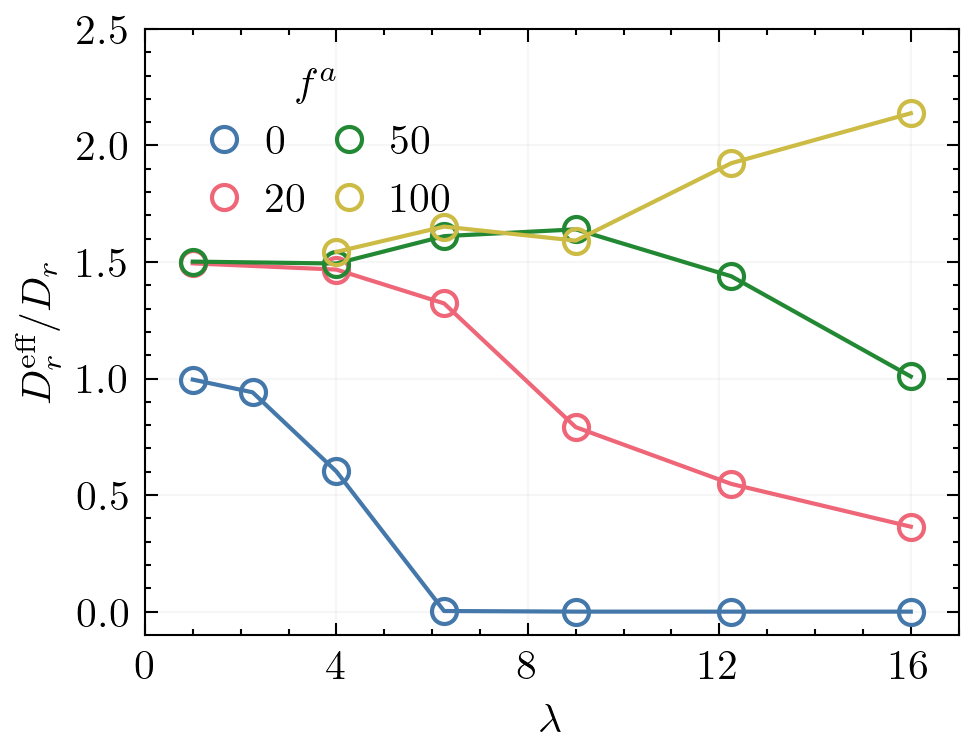}
    \caption{Effective rotational diffusion coefficient normalized by $D_r$ of freely rotating particles $D_r^{\text{eff}}/D_r$ as a function of the dipolar coupling $\lambda$ for active forces $f^a=$ 0, 20, 50, and 100. 
       }
    \label{fig:D_r}
\end{figure}
\subsection{Orientational dynamics}

We now discuss the orientational dynamics of dipolar active particles by investigating their orientational TACFs at various dipolar coupling strengths and activity levels. Fig.~\ref{fig:OCF} shows $C_e(t)$ at dipolar coupling strengths $\lambda = 4$ and $\lambda = 16$, and for active forces $f^{a}$ ranging from 0 to 200. Furthermore, for comparison purposes, the orientational  TACF  of a freely rotating particle, decaying as $\exp(-2D_rt^*)$ with $D_r=1$, is illustrated by dotted lines. In the passive limit ($f^{a}=0$), the decay of orientational TACF of dipolar particles slows down significantly with the increase of dipolar coupling strength as strong attractive dipolar torques constrain the rotational motion of particles.  Particularly in the strong coupling regime  of $\lambda=16$, the trapping of particles in a network of chain-like aggregates  results in an arrested  orientational dynamics and $C_e$ does not decay in the investigated time interval. As the active force increases, the orientational  TACFs decay faster at both coupling strengths.

For particles interacting with orientational torques, deviations from exponential form may occur. For passive systems, we observe deviations from exponential behavior for $\lambda \geq 4$. In such cases, $C_e(t)$
is found to be well-described by the stretched exponential function $\exp[-(t/\tau_r)^{\beta}]$, which models non-exponential relaxation behavior by a linear superposition of a continuous spectrum of exponential processes.
  The case of $\beta=1$ corresponds to a single exponential process, whereas a $\beta<1$ indicates a wider distribution of relaxation times, reflecting the effects of interparticle torques.
    From the characteristic decay time $\tau_r$ and stretching exponent $\beta$, the mean relaxation time can be computed as $\langle \tau_r \rangle = \tau_r/\beta \cdot \Gamma(1/\beta)$, where $\Gamma$ is the well-known gamma-function.   
    For passive systems   with $4 \le \lambda  < 16$  we found  stretching exponents $\beta$ ranging from $\sim 0.81$ to $\sim 0.28$ indicating a widening of distribution of relaxation times upon increase of coupling strengths.
    Interestingly upon switching on activity, we find that orientational TACFs of active systems
    always exhibit   stretching exponents $\beta\approx 1$, hence obeying  a single exponential decay processes. In any case, an effective rotational diffusion coefficient can be extracted from the relation $D_r^{eff}=1/2 \langle \tau_r \rangle$.

Fig.~\ref{fig:D_r} shows the extracted effective rotational diffusion coefficients $D_r^{eff}/D_r$  as a function of $\lambda$ for active forces $f^{a}=$ 0, 20, 50, and 100. In the passive limit ($f^{a}=0$), $D_r^{eff}\approx D_r $  for weakly dipolar interacting particles  and it globally decreases towards zero with increasing the coupling strength. For $\lambda \geq 6.25$, $D_r^{eff}$ approaches zero due to particles being trapped in large self-assembled structures highly limiting orientational fluctuations; see Fig.~\ref{fig:phase_diagram}. Upon switching on activity, a systematic enhancement of rotational diffusion is observed for all $\lambda$. This enhancement weakly depends on $f^{a}$ for $\lambda \leq 4$, where particles are mainly in monomeric form. For larger values of $\lambda$, we observe a notable dependence of $D_r^{eff}$ on both activity and dipolar coupling strength.  $D_r^{eff}$ decreases with $\lambda$
for  low active forces $f^a \le 50$, where the active systems preserve  the network structure, whereas it increases significantly with $\lambda$ for $f^a=100$, where active force dominates the dipolar attractive  interactions and leads to formation of an active gas. Remarkably, the effective rotational diffusion coefficient of strongly interacting dipolar active particles is approximately twice that of isolated active particles.
 Therefore, the interplay between activity and dipolar interactions  results in  distinct trends for the behavior of rotational diffusion depending on the coupling strength and the resulting non-equilibirum structure.
 \begin{figure}[h]
    \centering
     \begin{tikzpicture}
            \draw (0,0) node[inner sep=0]{\includegraphics[width=\linewidth]{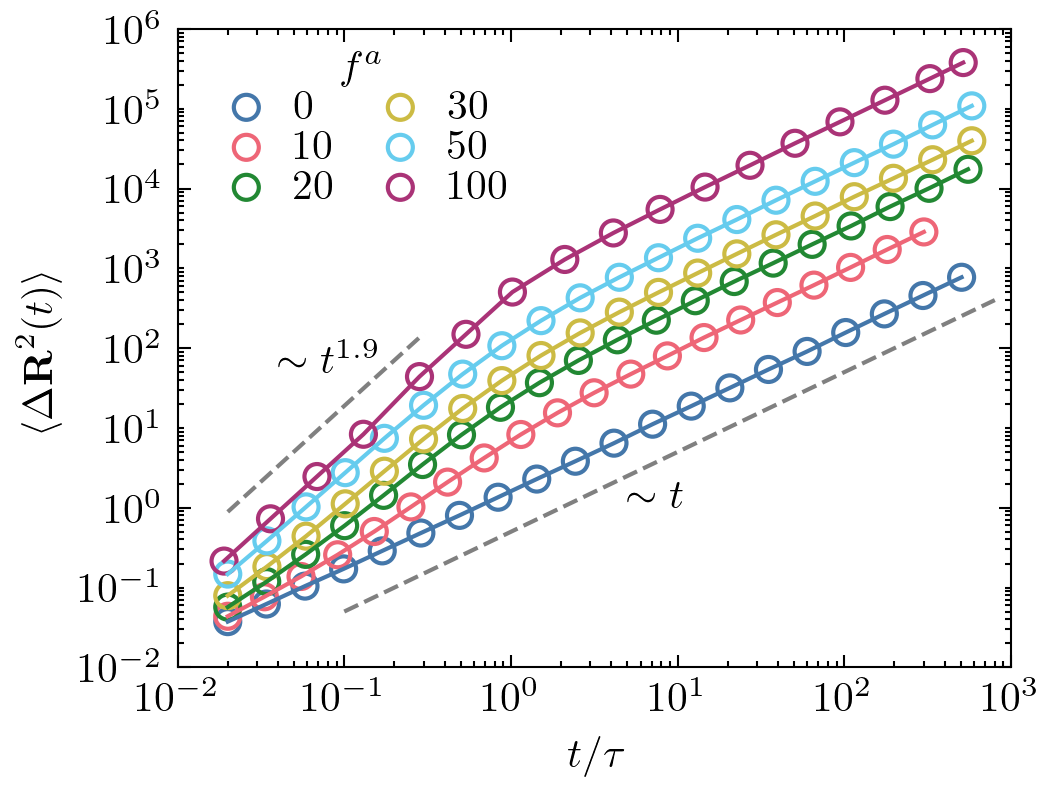}};
            \draw (-0.1,-6.6) node {\includegraphics[width=\linewidth]{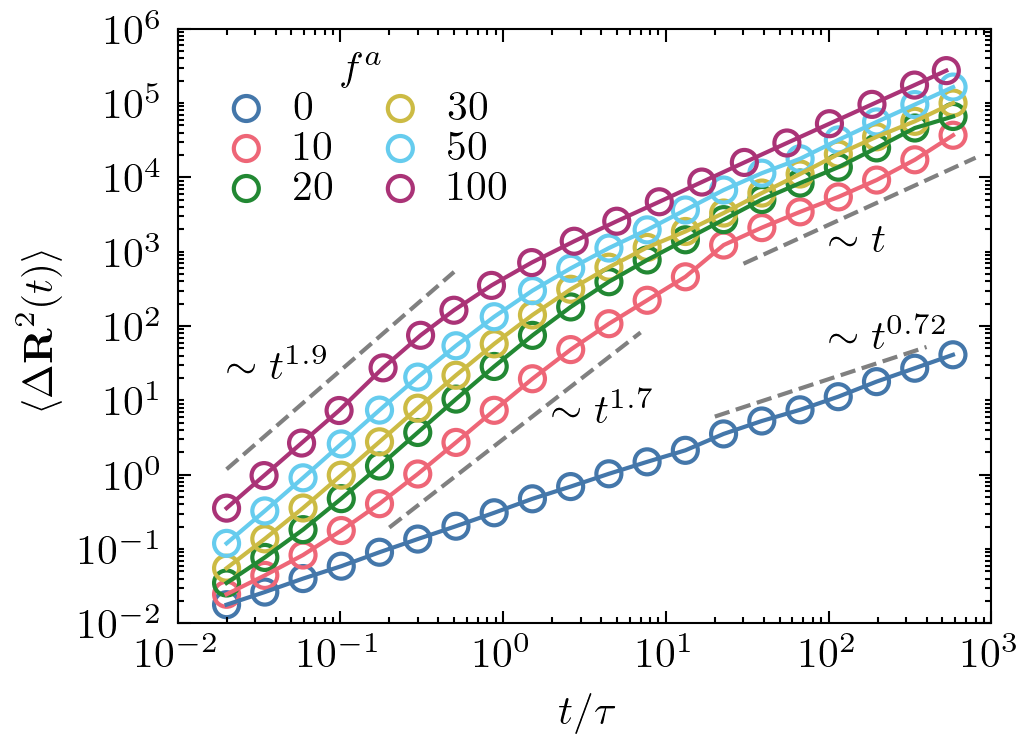}};
            \draw (-2.5,2.7) node {\textbf{(a)}};
            \draw (-2.5,-4) node {\textbf{(b)}};
    \end{tikzpicture}
    \caption{Mean-squared displacement  of   active particles with dipole-dipole interactions as   functions of time  for a) $\lambda=4$ and b) $\lambda=16$ at different self-propulsion speeds $U_0^*=0$,20,50,100 and 200.  }
    \label{fig:MSD}
\end{figure}

\begin{figure}[h]
    \centering
    \includegraphics[width=0.9\linewidth]{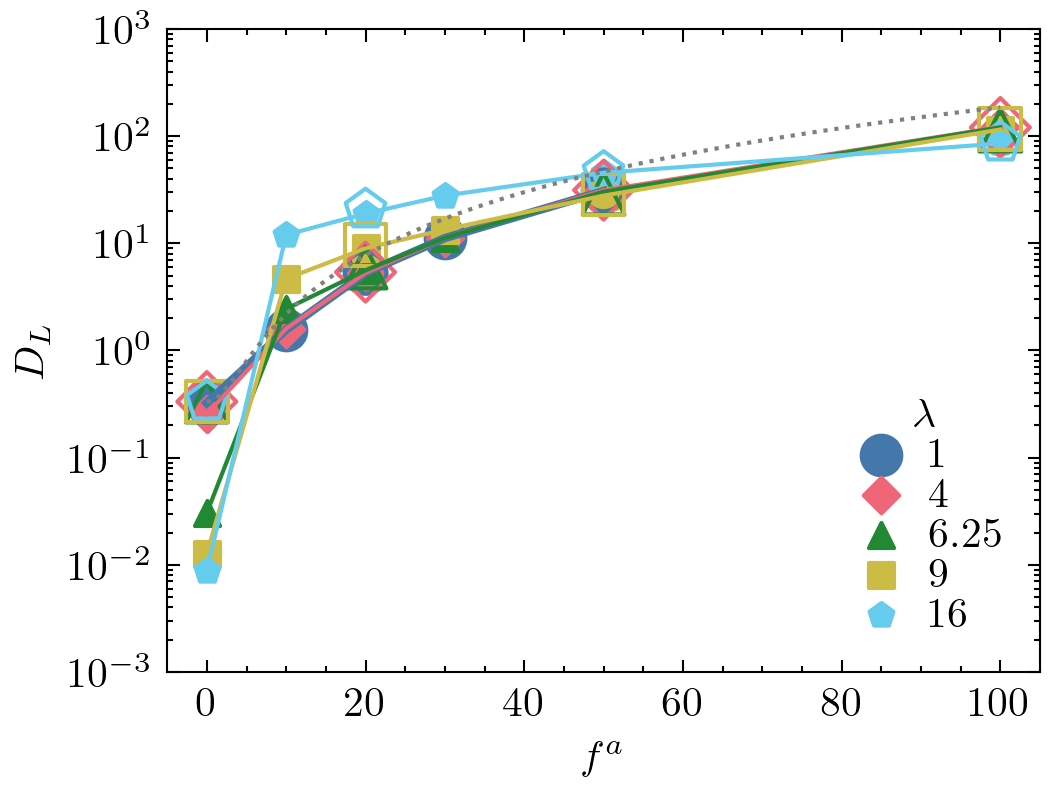}
    \caption{The long-time translational diffusion coefficients $D_L$  (close symbols) against $ f^a=3v^a $ for different dipolar coupling strengths $\lambda=4$,6.25, 9 and 16. The dotted line shows the long-time diffusion of non-interacting ABPs $D_L=D_t^*+\frac{{f^a}^2}{54 D_r}$, whereas   
    the open symbols show the effective long-time diffusion coefficient assuming active Brownian particles with effective rotational diffusion coefficient $D_r^{eff}$ shown in Fig.~\ref{fig:D_r} yielding $D_L^{eff}=D_t^*+\frac{{f^a}^2}{54D_r^{eff}}$.}
    \label{fig:D_L}
\end{figure}

\subsection{Translational dynamics }
Next, we explore the translational dynamics of active magnetic particles by computing their mean squared displacements (MSD) and their self-intermediate scattering functions. Fig.~\ref{fig:MSD} (a) and (b) present  MSD curves as functions of lag time for     dipolar coupling strengths $\lambda=4$ and 16 at  different active forces.  
At the lower coupling strength of $\lambda=4$, where particles are mainly in monomeric state, we observe a crossover from a ballistic regime at short times $t^*=t/\tau <1$ to a diffusive regime at long times for all $f^a$ values. At short lag times, the motion of  active magnetic particles can be roughly interpreted as moving straight with a constant speed of self-propulsion $v^a=f^a/3$.   
At the higher coupling strength of $\lambda=16$, in the passive limit, strongly interacting particles form a percolated network and we observe a subdiffusive behavior with MSD scaling as $\langle \Delta \vec{r}^2(t) \rangle \sim t^{0.72}$. The observed subdiffusive regime is a consequence of particles being trapped in the percolated network.  For active systems,  we observe a ballistic regime at very short times $t/\tau <0.1$, followed by a transient superdiffusive regime at intermediate times with an exponent smaller than 2 and a diffusive regime at long times $t/\tau > 10$.

In both low and strong coupling regimes, the slope of MSD at long times depends on the activity level.  We extract the long-time diffusion coefficient $D_L$ from linear fits of the MSD curves at large lag times from $\langle \Delta \vec{r}^2 (t^* \gg 1) \rangle  = 6D_{L}t^*$. Fig.~\ref{fig:D_L} presents extracted values of $D_L$  against $f^a$  for different values of dipolar coupling strengths.
We observe an increase in $D_L$ with active force across all cases. However, the extent of this dependence on $f^a$  varies with the strength of dipolar coupling.
Specifically, for the case of $\lambda=16$, the dependence of $D_L$ on $f^a$ is considerably weaker compared to that of non-interacting active Brownian particles, given by $D_L^{ABP}=D_t^*+\frac{{v^a}^2}{6D_r}=D_t^*+\frac{{f^a}^2}{54D_r}$, shown by the dotted line in Fig.~\ref{fig:D_L}. Interestingly, at low active forces, the long-time diffusion of the active gel is enhanced compared to non-interacting active particles, whereas for the active string fluid, $D_L$ is slightly smaller than $D_L$ of freely rotating active particles. To account for the effect of long-range dipolar interactions on slowing down of the long-time diffusion, we have also included in Fig.~\ref{fig:D_L} an effective  long-time translational diffusion, which is computed as
$D_L^{eff}=D_t^*+\frac{{f^a}^2}{54D_r^{eff}}$, where instead of the free particle rotational diffusion coefficient, we have used the effective rotational diffusion $D_r^{eff}$ obtained from orientational correlations, see Fig.~\ref{fig:D_r}. The calculated values of $D_L^{eff}$ are depicted by open symbols in Fig.~\ref{fig:D_L}, offering a reasonably accurate estimate of the diffusion coefficient $D_L$ for dipolar active particles. These results also serve to rationalize the deviations from enhanced diffusion of non-interacting active particles.
  The slowdown in rotational diffusion induced by strong dipole-dipole interactions leads to increase of persistence time of active particles which in turn results in enhanced  long-time translational diffusion. 
 %
 
\begin{figure}
    \centering
     \begin{tikzpicture}
            \draw (0,0) node[inner sep=0]{\includegraphics[width=\linewidth]{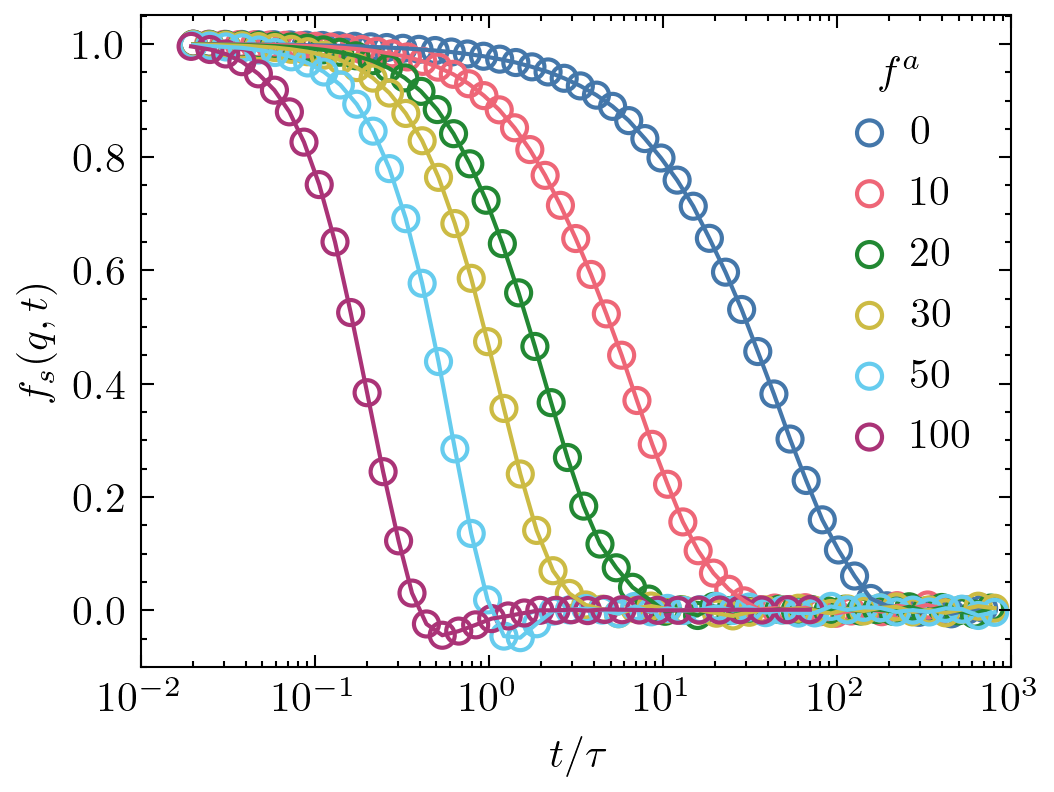}};
            \draw (-0.1,-6.6) node {\includegraphics[width=\linewidth]{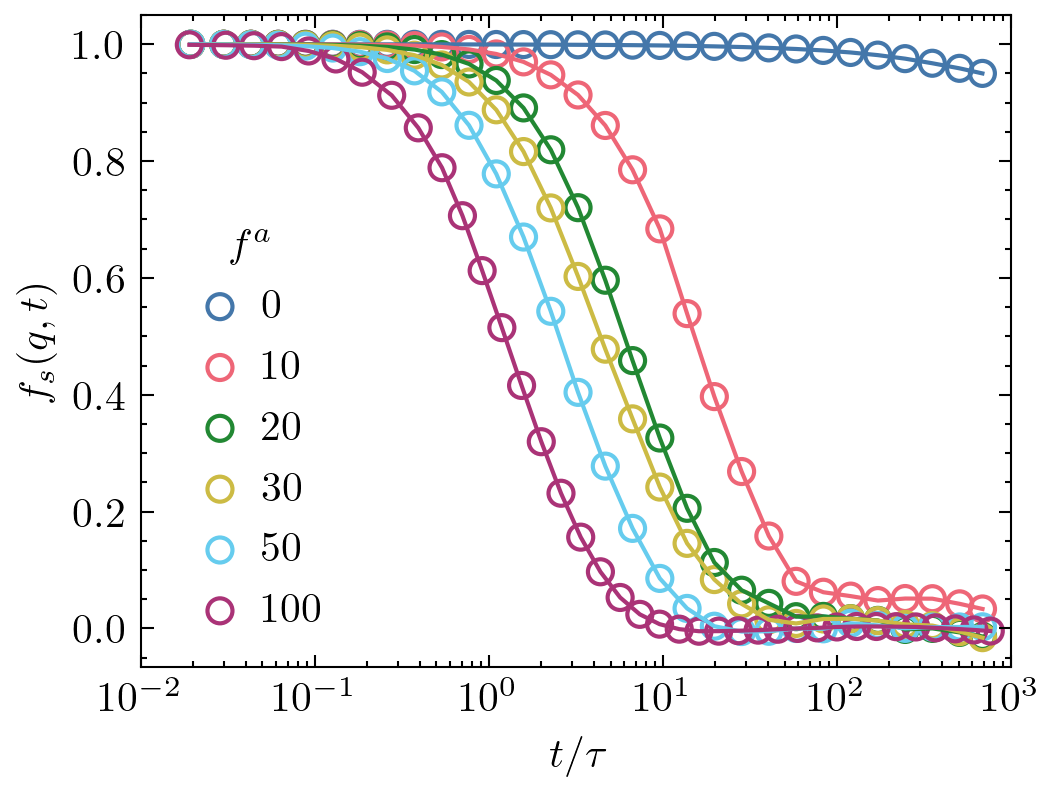}};
            \draw (-2.7,2.5) node {\textbf{(a)}};
            \draw (-2.7,-4.2) node {\textbf{(b)}};
    \end{tikzpicture}
    \caption{Self intermediate scattering function for a) $\lambda=4$ at $q\sigma=0.3$  and b) $\lambda=16$ at $q\sigma=0.08$ for different active forces  $f^a=0$,10,20,50 and 100.}
    \label{fig:SISF}
\end{figure}
 \begin{figure}
    \centering
\begin{tikzpicture}
            \draw (0,0) node[inner sep=0]{\includegraphics[width=\linewidth]{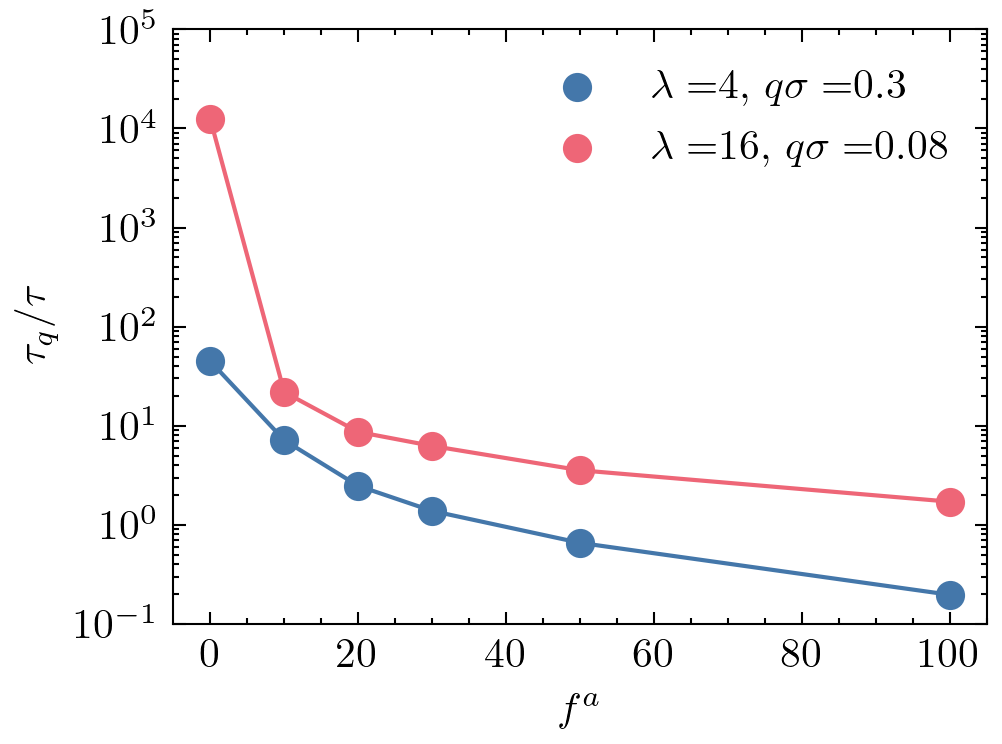}};
            \draw (-0.2,-6.) node {\includegraphics[width=\linewidth]{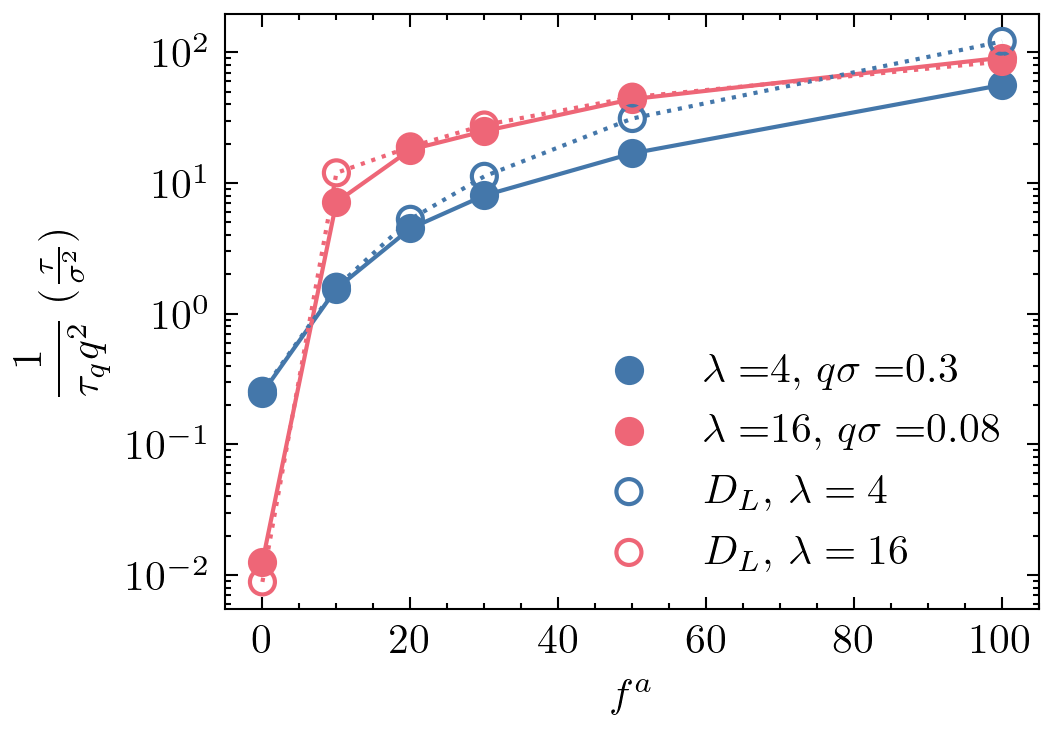}};
            \draw (-2.,2.2) node {\textbf{(a)}};
            \draw (-2.,-4.) node {\textbf{(b)}};
    \end{tikzpicture}
    
    \caption{(a) The dependence of relaxation times, $\tau_q/\tau$ on active force $f^a$ which are derived from intermediate scattering functions, $f_s(q,t)$ computed at $q\sigma=0.3$ and $q\sigma=0.08$ for   $\lambda=4$ and $16$, respectively. b)  The effective translational diffusion coefficient defined as $D_t^{eff}=1/(\tau_q q^2)$ versus $f^a$ compared to $D_L$ extracted from MSDs.
    }
    \label{fig:tau_q}
\end{figure}

Finally, we compute the self-intermediate scattering functions as defined in Eq.~\eqref{eq:F_s} for different activity levels for moderate and high coupling strengths. Fig.~\ref{fig:SISF} (a) and (b) present $f_s(q,t)$ as a function  of lag time for dipolar coupling strengths $\lambda=4$ and 16 at  different active forces at wavenumbers corresponding to the peaks of the structure factor of passive system counterparts for each dipolar coupling strength. In both cases, the increase of active force systematically leads to a faster decay of self-intermediate scattering function. The changes in decay of $f_s(q,t)$ with $f^a$ are especially remarkable at $\lambda=16$, where in the passive case the colloidal gel exhibits an arrested dynamics at the investigated time scales. However, a small active force $f^a=10$ already leads to a full decay of $f_s$ within $t/\tau=100$ despite the fact that the system maintains the percolated network structure. This reflects the role of dynamical rearrangements of network structures as demonstrated by a decrease of  mean bond lifetime with the activity level as demonstrated in Fig.~\ref{fig:BCF_fit}. 

 To quantify the dependence of relaxation of $f_s(q,t)$ on active force, we have fitted it with an exponential function of the form $\exp(-t/\tau_q)$ in the range $0\leq t\leq t_s$, where $f_s(q,t_s)=1/e$. For graphs that never reach $1/e$, we simply used the total simulation time as the range of our fitting. The extracted  relaxation times $\tau_q$ as functions of $f^a$  for $\lambda=4$ and 16 are presented in Fig.~\ref{fig:tau_q} (a). The drop of $\tau_q$ with $f^a$ is notably pronounced  for the high coupling regime with $\lambda=16$ where system forms an active gel. Using these relaxation times, we can derive an effective translational diffusion coefficient as $D_t^{eff}=1/(\tau_q q^2)$. The plot in Fig.~\ref{fig:tau_q}(b) illustrates both $D_t^{eff}$ and $D_L$ directly obtained from MSDs. We find  good agreement between the two diffusion coefficients, especially for the strong coupling strength of $\lambda=16$.

\section{Concluding remarks} \label{sec:conclusion}

This study presents an in-depth characterization of
the structural  and dynamical properties of low density ($\phi=0.01$) three-dimensional active particles with anisotropic dipole-dipole interactions obtained by Brownian dynamics simulations.  We investigated their non-equilibrium structure formation at a fixed density while varying the  dipolar coupling strength $\lambda$ and active force $f^a$. We find that the interplay between activity and long-range dipolar interactions results in the emergence of  two interesting dynamical self-assembled structures. The first one is an isotropic fluid comprised of active chains,  which occurs  at low to intermediate activity levels for sufficiently strong dipolar coupling strengths.
The second emerging structure  forms at very high dipolar coupling strengths and low  to moderate  activity levels and can be identified as an active gel   characterized by an interconnected percolated network of active particles


The active gel has a structure  similar to the  equilibrium gel found for passive dipolar spheres at low  temperatures~\cite{Sciortino_2012}. However, its configuration is more dynamic because of more frequent  rearrangements  within the network structure as evidenced by smaller mean bond lifetime  in active gels. Moreover,  active gels exhibit  significantly accelerated rotational and translational diffusion,  while retaining their interconnected network structures. Likewise, the observed active string fluids exhibit  enhanced dynamics compared to their passive counterparts. Our comprehensive investigation of the distribution of chain and ring lengths across various activity levels reveals that motility serves as a versatile tool for controlling the average size of aggregates or triggering a structural transition from a gel network to a string fluid.

In conclusion, our study offers fresh insights into the dynamic self-assembly behavior of active dipolar particles in three dimensions. We find that the interplay between long-range anisotropic interactions and three-dimensional activity fosters the formation of active fluids of strings and active gels. Magnetotactic bacteria exhibit a diverse range of sizes ($2<\sigma<10$ $\mu$m), dipole moments ($10^{-16} \le \mu \le 10^{-14}$ Am$^2$), and self-propulsion speeds ($10 <V^a < 100$ $\mu$ms$^{-1}$)~\cite{magnetotactic_bacteria_Bazylinski}. In dimensionless units and assuming a spherical shape, these ranges roughly translate to $0.1<\lambda<20$ and $5<f^a<200$. Consequently, our findings suggest the potential discovery of active strings and gels in collectives of magnetotactic bacteria exhibiting large dipolar coupling strengths and low self-propulsion speeds.

Our findings reveal that even in the absence of hydrodynamic interactions, intriguing active self-assembly behaviors emerge at moderate to high dipolar coupling strengths. However, it has been demonstrated that hydrodynamic contributions play a crucial role in dilute suspensions of weakly magnetic swimmers in external fields~\cite{Koessel_2019,Koessel_2020}. Hence, it will be pertinent in future research to investigate the interplay between hydrodynamic and magnetic dipolar interactions in collective structure formation. A promising avenue for future research involves exploring the effects of external fields, such as magnetic fields or shear deformation, on the self-assembly behavior of active dipolar particles. Especially, investigation of self-assembly response to shear deformation could unveil novel activity-induced rheological properties~\cite{Vinceti_19}.
Furthermore, an outstanding open question concerns the influence of dipolar interactions on non-equilibrium structure formation at higher densities. Specifically, it would be of interest to explore whether activity-induced orientational order emerges at moderate densities, below the threshold where passive dipolar particles exhibit orientational order~\cite{Weis_dipol_order_2006}, and whether a flocking transition similar to that observed in two-dimensional systems~\cite{dipolar_active_abp_2d_klapp} occurs.

\section*{Acknowledgements}
This work used the Dutch national e-infrastructure with the support of the
SURF Cooperative using grant no. EINF-5810.

\bibliography{magnetic.bib}
\appendix
\section*{Appendix}

Here, we provide additional plots to support the main article.
\begin{figure}[h]
    \centering
    \includegraphics[width=0.9\linewidth]{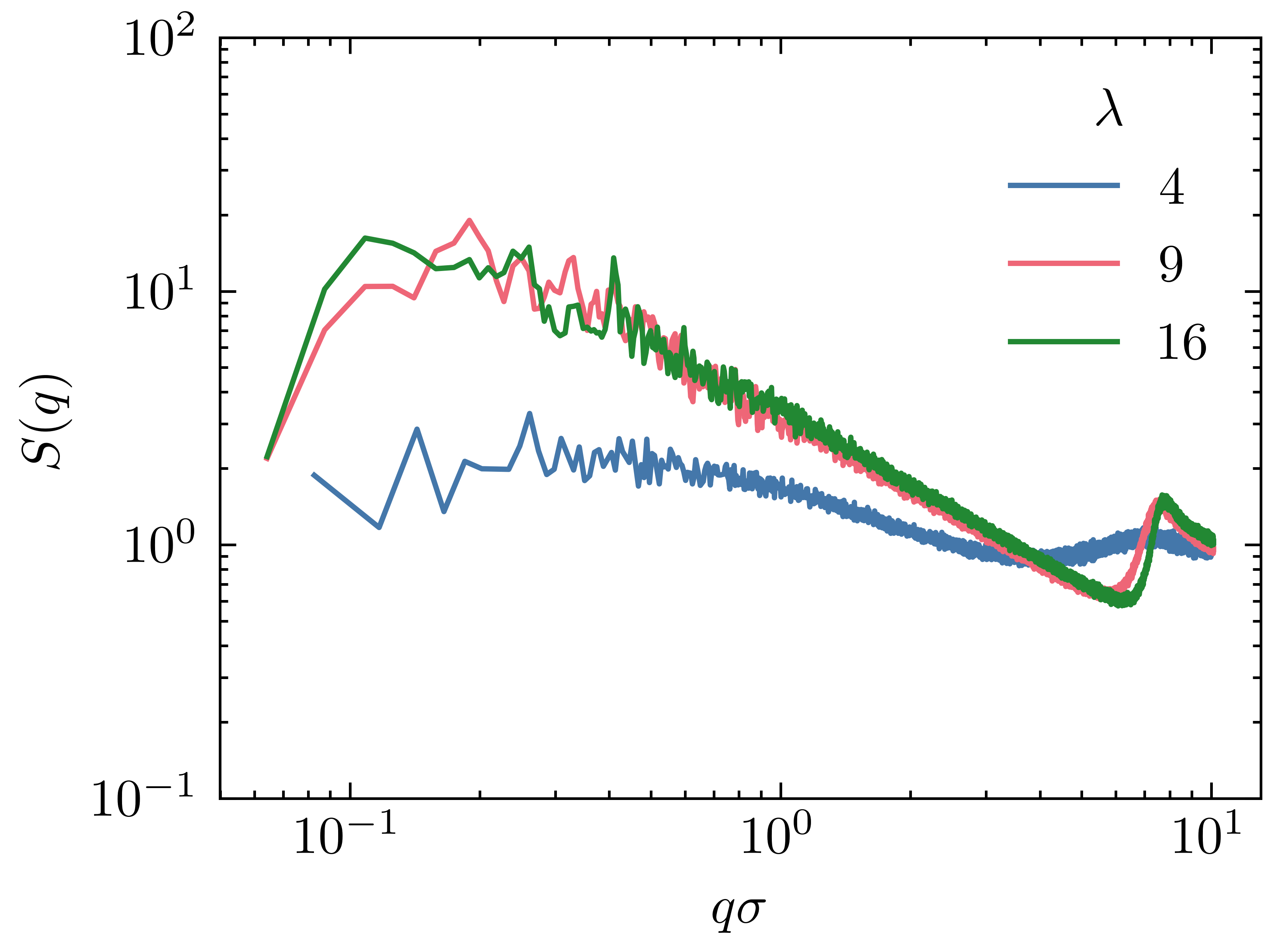}
    \caption{ Structure factor $S(q)$ of passive particles with different dipolar coupling strengths. }
    \label{fig:Sq_passive}
\end{figure}

\begin{figure}[h]
    \centering
    \includegraphics[width=0.9\linewidth]{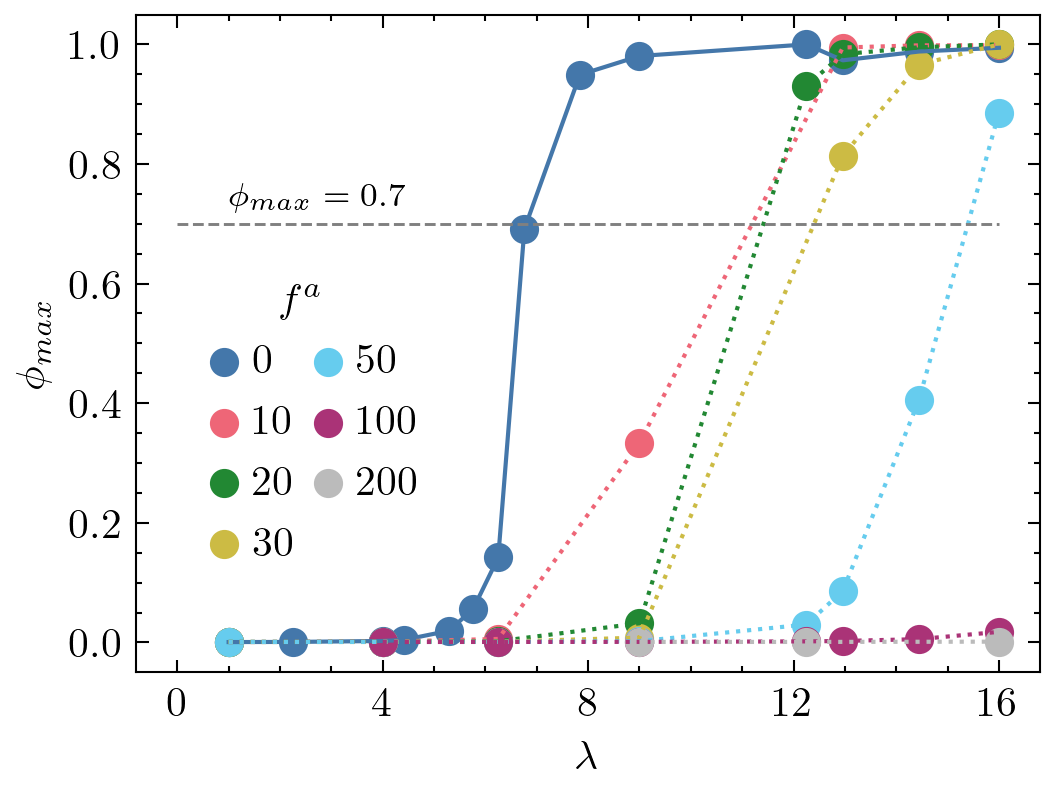}
    \caption{Mean fraction of particles in the largest cluster $\phi_{max}$ as a function of dipole-dipole coupling strength  $\lambda$ for different reduced active force $f^a$.}
    \label{fig:phi_max_lambda}
\end{figure}

\end{document}